\documentclass[12pt,a4paper]{article}

\usepackage{epsf}
\usepackage{latexsym,amssymb,euscript}
\usepackage[dvips]{graphicx}
\usepackage{amsmath}
\usepackage[nolists,nomarkers]{endfloat}

\newcommand{\beq}[1]{
\begin{equation}\label{#1}}
\newcommand{\eeq}{\end{equation}}
\newcommand{\bea}[1]{
\begin{eqnarray}\label{#1}}
\newcommand{\eea}{\end{eqnarray}}

\makeatletter
\def\appendix{\par\clearpage
  \setcounter{section}{0}
  \setcounter{subsection}{0}
  \@addtoreset{equation}{section}
  \def\@sectname{Appendix~}
  \def\theequation{\thesection.\arabic{equation}}
  \def\theequation{\thesection.\arabic{equation}}
  \def\thesection{\Alph{section}}}
\makeatother
\begin{document}
\begin{titlepage}

\begin{center}
{\LARGE \bf Mueller-Navelet small-cone jets at LHC in next-to-leading
BFKL}
\end{center}

\vskip 0.5cm

\centerline{F.~Caporale$^{1\dagger}$, D.Yu.~Ivanov$^{2\P}$,
B.~Murdaca$^{1\dagger}$ and A.~Papa$^{1\dagger}$}

\vskip .6cm

\centerline{${}^1$ {\sl Dipartimento di Fisica, Universit\`a della Calabria,}}
\centerline{\sl and Istituto Nazionale di Fisica Nucleare, Gruppo collegato di
Cosenza,}
\centerline{\sl I-87036 Arcavacata di Rende, Cosenza, Italy}

\vskip .2cm

\centerline{${}^2$ {\sl Sobolev Institute of Mathematics and
Novosibirsk State University,}}
\centerline{\sl 630090 Novosibirsk, Russia}

\vskip 2cm

\begin{abstract}
We consider within QCD collinear factorization the process $p+p\to {\rm jet}
+{\rm jet} +X$, where two forward high-$p_T$ jets are produced with a
large separation in rapidity $\Delta y$ (Mueller-Navelet jets).
In this case the (calculable) hard part of the reaction receives large
higher-order corrections $\sim \alpha^n_s (\Delta y)^n$, which can be
accounted for in the BFKL approach with next-to-leading logarithmic accuracy,
including contributions $\sim \alpha^n_s (\Delta y)^{n-1}$. We calculate
several observables related with this process, using the next-to-leading order
jet vertices, recently calculated in the approximation of small aperture of
the jet cone in the pseudorapidity-azimuthal angle plane.
\end{abstract}

%\vskip .5cm

$
\begin{array}{ll}
^{\dagger}\mbox{{\it e-mail address:}} &
\mbox{francesco.caporale, beatrice.murdaca, alessandro.papa \ @fis.unical.it}\\
^{\P}\mbox{{\it e-mail address:}} &
\mbox{d-ivanov@math.nsc.ru}\\
\end{array}
$

\end{titlepage}

\vfill \eject

\section{Introduction}

We consider the inclusive production at high energies of two forward
high-$k_T$ jets in proton-proton collisions,
\beq{process}
p(p_1) + p(p_2) \to {\rm jet}(k_{J_1}) +
{\rm jet}(k_{J_2})+ X \;,
\eeq
which are detected in the fragmentation regions of two colliding protons,
$p(p_1)$ and $p(p_2)$, and are separated by a large interval of rapidity
$\Delta y$, the so-called Mueller-Navelet (MN) process~\cite{Mueller:1986ey}.
This is considered as an important process for the manifestation of the
BFKL~\cite{BFKL} dynamics at hadron colliders, such as Tevatron and LHC.

\begin{figure}[t]
\centering
\includegraphics[scale=0.7]{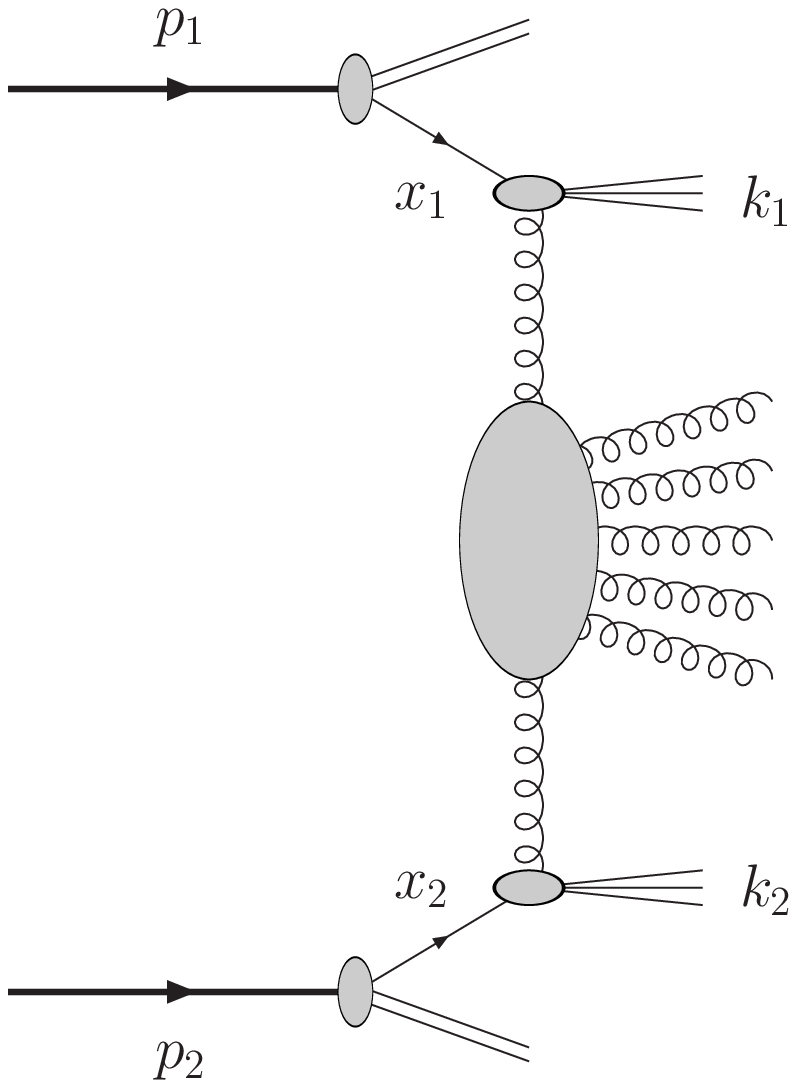}
\caption[]{%Schematic representation of the
Mueller-Navelet jet production process.}
\label{fig:MN}
\end{figure}

The theoretical description of this process is based on the QCD collinear
factorization. Neglecting higher-twist contributions (terms suppressed with
respect to the leading scaling asymptotic by additional inverse
powers of the hard scale), the process can be viewed as started by two hadrons
each emitting one parton, according to its parton distribution function (PDF),
with the subsequent partonic hard scattering, see Fig.~\ref{fig:MN}.
Collinear factorization allows to systematically resum the logarithms
of the hard scale, calculating the standard DGLAP evolution~\cite{DGLAP} of
the PDFs and the fixed-order radiative corrections to the parton scattering
cross section.

On the other side, in our kinematics at large squared center-of-mass
energy $\sqrt{s}$, when the rapidity gap between the two produced forward
jets is large, the BFKL resummation of energy logarithms comes into play,
since large logarithms of the energy compensate the small QCD coupling and
must be resummed to all orders of perturbation theory.

In comparison to the fixed-order DGLAP calculation, where an almost
back-to-back emission is expected, the BFKL calculation assumes more
emission of partons between the two jets and leads generically to a larger
cross-section and to a reduced azimuthal correlation between the detected
two forward jets.

At present the BFKL approach provides a general framework for the resummation
of energy logarithms in the leading logarithmic approximation (LLA), which
means resummation of all terms $(\alpha_s\ln(s))^n$, and in the
next-to-leading logarithmic approximation (NLA), which means resummation of
all terms $\alpha_s(\alpha_s\ln(s))^n$. Such resummation is
process-independent and is encoded in the Green's function for the
interaction of two Reggeized gluons. The Green's function is determined
through the BFKL equation, which is an iterative integral equation, whose
kernel is known at the next-to-leading order (NLO) both for forward
scattering ({\it i.e.} for $t=0$ and color singlet in the
$t$-channel)~\cite{FL98,CC98} and for any fixed (not growing with energy)
momentum transfer $t$ and any possible two-gluon color state in the
$t$-channel~\cite{Fadin:1998jv,FG00,FF05}.

The process-dependent part of the information needed for constructing the
cross section for the production of Mueller-Navelet jets is contained
in the impact factors for the transition from the colliding parton to the
forward jet (the so called ``jet vertex''). Mueller-Navelet jet vertices were
calculated with NLO accuracy in~\cite{Bartels:2002yj}.
The results of~\cite{Bartels:2002yj} were then used
in~\cite{Colferai:2010wu,Ducloue:2012bm} for a numerical estimation in the NLA
of the cross section for Mueller-Navelet jets at LHC and for the analysis of
the azimuthal correlation of the produced jets. This numerical analysis
followed previous ones~\cite{Vera:2007kn,Marquet:2007xx} based on the
inclusion of NLO effects only in the Green's functions. Recently we performed
a new calculation~\cite{Caporale:2011cc} of the jet impact factor, confirming
the results of~\cite{Bartels:2002yj}.

Although NLO jet vertices were obtained for a general jet algorithm, the
implementation of these results into the cross section and the other
Mueller-Navelet jet observables calculation requires a rather complicated
numerical evaluation, see~\cite{Colferai:2010wu}.
Recently the NLO impact factor for the production of forward jets was
calculated by two of us~\cite{Ivanov:2012ms} in the ``small-cone''
approximation (SCA)~\cite{Furman:1981kf,Aversa}, {\it i.e.} for small jet cone
aperture in the rapidity-azimuthal angle plane. The use of the SCA allowed
to get a simple analytic result for the jet vertices in the so-called
$(\nu,n)$-representation (the jet vertices projected on the eigenfunctions
of the BFKL kernel) which can be  easily implemented in numerical calculations.
It is the aim of the present paper to obtain predictions for the
Mueller-Navelet jet process cross section and for the azimuthal angle
decorrelation observables in the SCA using the results of~\cite{Ivanov:2012ms}.

Before we proceed, let us further comment on the theoretical input behind 
our approach. Our basic assumption is to neglect entirely all contributions 
(higher twists) that are suppressed by additional inverse powers of the jet 
transverse momenta. Due to this, we work in collinear factorization and 
express our results in terms of the usual collinear PDFs and the hard part, 
{\it i.e.} the cross section for the inclusive production of two jets, 
initiated by two partons collinear to the incident protons.

The hard part is calculated within the BFKL NLA approach, which means that we 
neglect contributions that are suppressed by additional inverse powers of the
energy and resum to all orders in a model-independent way the leading and 
the first subleading energy logarithms.

Another approach, widely used in small-$x$ phenomenology, is based on the 
high-energy, or $k_T$-factorization~\cite{Catani93}. In this case cross 
sections are expressed in terms of the unintegrated gluon density, which 
depends on both the longitudinal $x$ and the transverse $k_T$-part of the
gluon momentum. The $k_T$-factorization method allows also to resum the 
leading (and potentially first nonleading) logarithms of the energy, and what 
is more important, it allows to take into account some higher-twist 
contributions, which could play a significant role in the description of 
small-$x$ processes.    

In the Mueller-Navelet kinematics, each of the two jets is produced in the 
fragmentation region of one of the two incident protons, which ensures 
that the longitudinal momentum fractions of both partons that initiate the 
hard scattering are not small.   

In other words, despite its diffractive nature, the Mueller-Navelet process 
does not belong to the class of small-$x$ reactions. Therefore the collinear 
factorization seems to be an adequate tool here. Another interesting 
possibility is to study processes where there is only one forward jet 
associated with some hard final state $X$ produced at central rapidities 
(say, $X$ includes one hard central jet). Such processes are initiated by 
small-$x$ gluons, therefore it is natural to describe them, contrary to the 
Mueller-Navelet process, using the $k_T$-factorization method, see for 
instance~\cite{Deak09}.

The paper is organized as follows: in the next Section we discuss the
kinematics and recall the formulae for the Mueller-Navelet jet process
cross section; in Section~3 we present our results; Section~4 contains the
discussion of results and our conclusions.

\section{Mueller-Navelet jet cross section}

It is convenient to define the Sudakov decomposition for the momenta of the
jets; one has
\begin{equation}
k_{J_1}= x_{J_1} p_1+ \frac{\vec k^2_{J_1}}{x_{J_1} s}p_2+k_{J_1 \, \perp} \ ,
\quad\quad\quad k_{J_1\, \perp}^2=-\vec k^2_{J_1} \ ,
\end{equation}
\begin{equation}
k_{J_2}= x_{J_2} p_2+ \frac{\vec k^2_{J_2}}{x_{J_2} s}p_1+k_{J_2 \, \perp} \ ,
\quad\quad\quad k_{J_2\, \perp}^2=-\vec k^2_{J_2} \;.
\end{equation}
Here as power-suppressed correction we neglect the proton mass,
$p_1^2=p_2^2=0$, therefore $p_1$ and $p_2$ are taken as Sudakov vectors
satisfying  $2(p_1 p_2)=s$.

For the forward jets the longitudinal fractions $x_{J_{1,2}}\sim {\cal O}(1)$
are related to the jet rapidities in the center-of-mass system by
\[
y_1=\frac{1}{2}\ln\frac{x_{J_1}^2 s}{\vec k_{J_1}^2}\;,\quad\quad
dy_1=\frac{dx_{J_1}}{x_{J_1}}\;,
\]
\[
y_2=-\frac{1}{2}\ln\frac{x_{J_2}^2 s}{\vec k_{J_2}^2}\;,\quad\quad
dy_2=-\frac{dx_{J_2}}{x_{J_2}}\;,
\]
so that the rapidity gap between the two jets is given by
\beq{Y}
\Delta y \equiv Y=\ln\frac{x_{J_1} x_{J_2} s}{|\vec k_{J_1}||\vec k_{J_2}|}\;.
\eeq
For the Mueller-Navelet jet process, the rapidity gap $Y$ is to be taken much
larger than unity, thus implying the kinematics
\beq{kin}
s\gg \vec k^2_{J_{1,2}}\gg \Lambda^2_{QCD} \, ,
\eeq
and the transverse momenta of the jets are assumed to be of similar order of
magnitude, $\vec k^2_{J_1}\sim \vec k^2_{J_2}$.

In QCD collinear factorization the cross section of the process~(\ref{process})
reads
\begin{equation}
\frac{d\sigma}{dx_{J_1}dx_{J_2}d^2 k_{J_1}d^2 k_{J_2}}
=\sum_{i,j=q,\bar q,g}\int\limits^1_0 dx_1 \int\limits^1_0 dx_2\,f_i(x_1,\mu_F)
f_j(x_2,\mu_F) \frac{d\hat \sigma_{i,j}(x_1 x_2 s,\mu_F)}{dx_{J_1}dx_{J_2}d^2
 k_{J_1}d^2 k_{J_2}}\;,
\label{ff}
\end{equation}
where the $i,j$ indices specify the parton types (quarks $q=u,d,s,c,b$;
antiquarks $\bar q=\bar u,\bar d,\bar s,\bar c,\bar b$; or gluon $g$),
$f_i(x,\mu_F)$ denotes the initial proton PDFs, the longitudinal
fractions of the partons involved in the hard subprocess are $x_{1,2}$,
as shown in Fig.~\ref{fig:MN}, $\mu_F$ is the factorization scale,
$d\hat \sigma_{i,j}(x_1 x_2 s,\mu_F)$ is the partonic cross section for the
production of jets and $\hat s=x_1 x_2 s$ is the squared center-of-mass
energy of the parton-parton collision subprocess. We use the
$\overline{\mbox{MS}}$ scheme for the ultraviolet and collinear factorizations.

At lowest order each jet is generated by a single parton having high
transverse momentum and the partonic subprocess is given by an elementary
two-to-two scattering. In the discussed Mueller-Navelet kinematics the
higher-order contributions to the partonic cross section have to be resummed
using BFKL approach. In the BFKL approach~\cite{BFKL}, the cross section of
the hard subprocess reads
\beq{hard}
\frac{d\hat \sigma_{i,j}(x_1 x_2 s)}{dx_{J_1}dx_{J_2}d^2
k_{J_1}d^2 k_{J_2}}=\frac{1}{(2\pi)^2}\int\frac{d^2 q_1}{\vec
q_1^{\,\, 2}} V_i(\vec q_1,s_0,x_1;\vec k_{J_1},x_{J_1})
\int\frac{d^2 q_2}{\vec q_2^{\,\,2}}V_j(-\vec q_2,s_0,x_2;
\vec k_{J_2},x_{J_2})
\eeq
\[
\times \int\limits^{\delta +i\infty}_{\delta
-i\infty}\frac{d\omega}{2\pi i}\left(\frac{x_1 x_2 s}{s_0}\right)^\omega
G_\omega (\vec q_1, \vec q_2)\, .
\]
This representation for the cross section is valid with NLA accuracy.
Here $V_i(\vec q_1,s_0,x_1;\vec k_{J_1},x_{J_1})$ and
$V_j(-\vec q_2,s_0,x_2;\vec k_{J_2},x_{J_2})$ are the jet vertices
(impact factors) describing the transitions ${\rm parton \ } i\ (x_1 p_1) \to
{\rm jet}\ (k_{J_1})$ and ${\rm parton \ } j \ (x_2p_2) \to {\rm jet}\
(k_{J_2})$, in the scattering off a Reggeized gluon with transverse momentum
$\vec q_1$ and $\vec q_2$, respectively. The artificial scale $s_0$ is
introduced in the BFKL approach to perform the Mellin transform from the
$s$-space to the complex angular momentum plane and cancels in the full
expression for the cross section with the NLA accuracy.
%They have been
%calculated in Refs.~\cite{Colferai:2010wu,Caporale:2011cc} for arbitrary
%infrared-safe jet selection function and in Ref.~\cite{IP12} in the
%approximation of small aperture of the jet cone in the
%pseudorapidity-azimuthal angle plane~\cite{Furman:1981kf,Aversa}.
The Green's function in~(\ref{hard}) obeys the BFKL equation
\beq{Green}
\delta^2(\vec q_1-\vec q_2)=\omega \, G_\omega (\vec q_1, \vec q_2)-
\int d^2 q \, K(\vec q_1,\vec q)\, G_\omega (\vec q, \vec q_2) \;,
\eeq
where $K(\vec q_1,\vec q_2)$ is the BFKL kernel.

In what follows we proceed along the lines similar to ones used in
Ref.~\cite{mesons}. It is convenient to work in the transverse momentum
representation, defined by
\beq{transv}
\hat{\vec q}\: |\vec q_i\rangle = \vec q_i|\vec q_i\rangle\;,
\eeq
\beq{norm}
\langle\vec q_1|\vec q_2\rangle =\delta^{(2)}(\vec q_1 - \vec q_2) \;,
\hspace{2cm}
\langle A|B\rangle =
\langle A|\vec k\rangle\langle\vec k|B\rangle =
\int d^2k A(\vec k)B(\vec k)\;;
\eeq
the kernel of the operator $\hat K$ is
\beq{kernel-op}
K(\vec q_2, \vec q_1) = \langle\vec q_2| \hat K |\vec q_1\rangle
\eeq
and the equation for the Green's function reads
\beq{Groper}
\hat 1=(\omega-\hat K)\hat G_\omega\;,
\eeq
its solution being
\beq{Groper1}
\hat G_\omega=(\omega-\hat K)^{-1} \, .
\eeq

The kernel is given as an expansion in the strong coupling,
\beq{kern}
\hat K=\bar \alpha_s \hat K^0 + \bar \alpha_s^2 \hat K^1\;,
\eeq
where
\beq{baral}
{\bar \alpha_s}=\frac{\alpha_s N_c}{\pi}
\eeq
and $N_c$ is the number of colors. In Eq.~(\ref{kern}) $\hat K^0$ is the
BFKL kernel in the LLA, $\hat K^1$ represents the NLA correction.

To determine the partonic cross section with NLA accuracy we need an
approximate solution of Eq.~(\ref{Groper1}). With the required accuracy this
solution is
\beq{exp}
\hat G_\omega=(\omega-\bar \alpha_s\hat K^0)^{-1}+
(\omega-\bar \alpha_s\hat K^0)^{-1}\left(\bar \alpha_s^2 \hat K^1\right)
(\omega-\bar \alpha_s \hat
K^0)^{-1}+ {\cal O}\left[\left(\bar \alpha_s^2 \hat K^1\right)^2\right]
\, .
\eeq

The basis of eigenfunctions of the LLA kernel,
\beq{KLLA}
\hat K^0 |n,\nu\rangle = \chi(n,\nu)|n,\nu\rangle \, , \;\;\;\;\;\;\;\;\;\;
\chi (n,\nu)=2\psi(1)-\psi\left(\frac{n}{2}+\frac{1}{2}+i\nu\right)
-\psi\left(\frac{n}{2}+\frac{1}{2}-i\nu\right)\, ,
\eeq
is given by the following set of functions:
\beq{nuLLA}
\langle\vec q\, |n,\nu\rangle =\frac{1}{\pi\sqrt{2}}
\left(\vec q^{\,\, 2}\right)^{i\nu-\frac{1}{2}}e^{in\phi} \;,
%\;\;\;\;\; \cos\phi \equiv q_x\;,
\eeq
here $\phi$ is the azimuthal angle of the vector $\vec q$ counted from
some fixed direction in the transverse space, $\cos\phi \equiv q_x/|\vec q\,|$.
Then, the orthonormality  condition takes the form
\beq{ort}
\langle n',\nu^\prime | n,\nu\rangle =\int \frac{d^2 q}
{2 \pi^2 }\left(\vec q^{\,\, 2}\right)^{i\nu-i\nu^\prime -1}
e^{i(n-n')\phi}=\delta(\nu-\nu^\prime)\, \delta_{nn'}\, .
\eeq
The action of the full NLA BFKL kernel on these functions may be expressed
as follows:
\bea{Konnu}
\hat K|n,\nu\rangle &=&
\bar \alpha_s(\mu_R) \chi(n,\nu)|n,\nu\rangle
 +\bar \alpha_s^2(\mu_R)\left(\chi^{(1)}(n,\nu)
+\frac{\beta_0}{4N_c}\chi(n,\nu)\ln(\mu^2_R)\right)|n,\nu\rangle
\nonumber \\
&+& \bar
\alpha_s^2(\mu_R)\frac{\beta_0}{4N_c}\chi(n,\nu)
\left(i\frac{\partial}{\partial \nu}\right)|n,\nu\rangle \;,
\eea
where $\mu_R$ is the renormalization scale of the QCD coupling, the first
term represents the action of LLA kernel, while the second and the third ones
stand for the diagonal and the non-diagonal parts of the NLA kernel and we
have used
\beq{beta00}
\beta_0=\frac{11 N_c}{3}-\frac{2 n_f}{3}\;,
\eeq
where $n_f$ is the number of active quark flavors.

The function $\chi^{(1)}(n,\nu)$, calculated in~\cite{Kotikov:2000pm} (see
also~\cite{Kotikov:2000pm2}), is conveniently represented in the form
\beq{ch11}
\chi^{(1)}(n,\nu)=-\frac{\beta_0}{8\, N_c}\left(\chi^2(n,\nu)-\frac{10}{3}
\chi(n,\nu)-i\chi^\prime(n,\nu)\right) + {\bar \chi}(n,\nu)\, ,
\eeq
where
\beq{chibar}
\bar \chi(n,\nu)\,=\,-\frac{1}{4}\left[\frac{\pi^2-4}{3}\chi(n,\nu)
-6\zeta(3)-\chi^{\prime\prime}(n,\nu) +\,2\,\phi(n,\nu)+\,2\,\phi(n,-\nu)
\right.
\eeq
\[
+ \left.
\frac{\pi^2\sinh(\pi\nu)}{2\,\nu\, \cosh^2(\pi\nu)}
\left(
\left(3+\left(1+\frac{n_f}{N_c^3}\right)\frac{11+12\nu^2}{16(1+\nu^2)}\right)
\delta_{n0}
-\left(1+\frac{n_f}{N_c^3}\right)\frac{1+4\nu^2}{32(1+\nu^2)}\delta_{n2}
\right)\right] \, ,
\]
\beq{phi}
\phi(n,\nu)\,=\,-\int\limits_0^1dx\,\frac{x^{-1/2+i\nu+n/2}}{1+x}
\left[\frac{1}{2}\left(\psi'\left(\frac{n+1}{2}\right)-\zeta(2)\right)
+\mbox{Li}_2(x)+\mbox{Li}_2(-x) \right.
\eeq
\[
\left. +\ln x \left(\psi(n+1)-\psi(1)+\ln(1+x)+\sum_{k=1}^\infty\frac{(-x)^k}
{k+n}\right)+\sum_{k=1}^\infty\frac{x^k}{(k+n)^2}(1-(-1)^k)\right]
\]
\[
=\sum_{k=0}^\infty\frac{(-1)^{k+1}}{k+(n+1)/2+i\nu}\left[\psi'(k+n+1)
-\psi'(k+1)+(-1)^{k+1}(\beta'(k+n+1)+\beta'(k+1))\right.
\]
\[
\left.
-\frac{1}{k+(n+1)/2+i\nu}(\psi(k+n+1)-\psi(k+1))\right] \, ,
\]
\[
\beta'(z)=\frac{1}{4}\left[\psi'\left(\frac{z+1}{2}\right)
-\psi'\left(\frac{z}{2}\right)\right]\;, \;\;\;\;\;
\mbox{Li}_2(x)=-\int\limits_0^xdt\,\frac{\ln(1-t)}{t} \, .
\]
Here and below $\chi^\prime(n,\nu)=d\chi(n,\nu)/d\nu$ and
$\chi^{\prime\prime}(n,\nu)=d^2\chi(n,\nu)/d^2\nu$.

For the quark and the gluon jet vertices in~(\ref{hard}) the projection onto
the eigenfunctions of LO BFKL kernel, {\it i.e.} the transfer to the
$(\nu,n)$-representation, is done as follows:
\bea{nu_rep}
&&
\frac{V(\vec q_1)}{\vec q_1^{\,\, 2}}=\sum^{+\infty}_{n=-\infty}
\int\limits^{+\infty}_{-\infty}
d\nu \, \Phi_1(\nu,n)\langle n,\nu| \vec q_1\rangle\, , \quad
\frac{V(-\vec q_2)}{\vec q_2^{\,\, 2}}=\sum^{+\infty}_{n=-\infty}
\int\limits^{+\infty}_{-\infty} d\nu \, \Phi_2(\nu,n)
\langle \vec q_2 |n,\nu \rangle \, ,
\nonumber\\
&&
\Phi_1(\nu,n)=\int d^2 q_1 \,\frac{V(\vec q_1)}{\vec q_1^{\,\, 2}}
\frac{1}{\pi \sqrt{2}} \left(\vec q_1^{\,\, 2}\right)^{i\nu-\frac{1}{2}}
e^{i n \phi_1}\;, \nonumber \\
&&
\Phi_2(\nu,n)=\int d^2 q_2 \,\frac{V(-\vec q_2)}{\vec q_2^{\,\, 2}}
\frac{1}{\pi \sqrt{2}} \left(\vec q_2^{\,\, 2}\right)^{-i\nu-\frac{1}{2}}
e^{-i n \phi_2}\;.
\eea
The vertices can be represented as an expansion in $\alpha_s$,
\beq{vertex-exp}
\Phi_{1,2}(n,\nu)=\alpha_s(\mu_R) v_{1,2}(n,\nu)+ \alpha_s^2(\mu_R)
v_{1,2}^{(1)}(n,\nu) \, .
\eeq
In Eqs.~(\ref{nu_rep}) and~(\ref{vertex-exp}) we suppressed for brevity the
partonic indices $i,j$  and the other arguments in $v_{1,2}$.
The explicit forms of LLA and NLA jet vertices in the $(\nu,n)$-representation
both for the quark and gluon cases can be found in~\cite{Ivanov:2012ms}.
In particular for the LLA quark vertices one has
\bea{v1}
&&
v^q_1(n,\nu)=2\sqrt{\frac{C_F}{C_A}}
(\vec k_{J_1}^2)^{i\nu-3/2}\,e^{i n \phi_{J_1}}\delta(x_{J_1}-x_1) \;,
\nonumber \\
&&
v^q_2(n,\nu)=2\sqrt{\frac{C_F}{C_A}}
(\vec k_{J_2}^2)^{-i\nu-3/2}\,e^{-i n (\phi_{J_2}+\pi)}\delta(x_{J_2}-x_2) \;,
\eea
where $C_A=N_c$, $C_F=(N_c^2-1)/2N_c$, the angle $\phi_{J_2}+\pi$ in the last
equation appears due to the fact that the Reggeon momentum which enters the
second vertex is $-\vec q_2$. Note that in LLA vertices the partonic and
the jet longitudinal momentum fractions coincide.

The partonic cross section can be written with NLA accuracy as follows
\bea{NLA-cross-sect}
&&
\frac{d\hat \sigma(x_1 x_2 s)}{dx_{J_1}dx_{J_2}d^2 k_{J_1}d^2 k_{J_2}}
=\frac{1}{(2\pi)^2}
\sum^{+\infty}_{n=-\infty}\int\limits^{+\infty}_{-\infty}
d\nu \,\left(\frac{x_{1} x_{2} s}{s_0}
\right)^{\bar \alpha_s(\mu_R)\chi(n,\nu)}
\nonumber \\
&&
\times \alpha_s^2(\mu_R) v_1(n,\nu)v_2(n,\nu)\left[
1+\alpha_s(\mu_R)\left(\frac{v_1^{(1)}(n,\nu)}{v_1(n,\nu)}
+\frac{v_2^{(1)}(n,\nu)}{v_2(n,\nu)}\right)
\right.
\nonumber \\
&&
+\bar \alpha_s^2(\mu_R)\ln\left(\frac{x_{1}x_{2} s}{s_0}\right)
\left(
\bar \chi(n,\nu)+ \frac{\beta_0}{8 N_c} \chi(n,\nu)
\left[
-\chi(n,\nu) + \frac{10}{3} \right.\right.\nonumber \\
&&
\left.\left.\left.
+ i\frac{d\ln\left( \frac{v_1(n,\nu)}{v_2(n,\nu)}\right)}{d\nu}
 +2\ln \mu_R^2\right]
\right)\right]\, .
\eea
For the subsequent calculation it is convenient to make the substitution
\beq{factors}
\left(\frac{x_{1} x_{2} s}{s_0}
\right)^{\bar \alpha_s(\mu_R)\chi(n,\nu)}=\left(\frac{x_{J_1} x_{J_2} s}{s_0}
\right)^{\bar \alpha_s(\mu_R)\chi(n,\nu)}\left(\frac{x_{1}}{x_{J_1}}
\right)^{\bar \alpha_s(\mu_R)\chi(n,\nu)}\left(\frac{x_{2} }{x_{J_2}}
\right)^{\bar \alpha_s(\mu_R)\chi(n,\nu)} \, ,
\eeq
and to assign the last two factors in the r.h.s. to the corresponding
jet vertices. This procedure affects only the NLA parts of the jet vertices,
since for the LLA vertices $x_i=x_{J_i}$. Also with NLA accuracy, one can
make in~(\ref{NLA-cross-sect}) the replacement
\beq{}
\ln\left(\frac{x_{1}x_{2} s}{s_0}\right)\to \ln\left(\frac{x_{J_1}x_{J_2} s}
{s_0}\right)\, .
\eeq
This procedure allows to perform in the MN-jet cross section first the
integration over partonic momentum fractions, before taking the sum over
$n$ and the integration over $\nu$; it allows also to consider together the
contributions of quarks and gluons to the jet vertices.

The differential cross section has the form
\beq{}
\frac{d\sigma}
{dy_{J_1}dy_{J_2}\, d|\vec k_{J_1}| \, d|\vec k_{J_2}|
d\phi_{J_1} d\phi_{J_2}}
=\frac{1}{(2\pi)^2}\left[{\cal C}_0+\sum_{n=1}^\infty  2\cos (n\phi )\,
{\cal C}_n\right]\, ,
\eeq
where $\phi=\phi_{J_1}-\phi_{J_2}-\pi$, and
\beq{Cm}
{\cal C}_m = \int_0^{2\pi}d\phi_{J_1}\int_0^{2\pi}d\phi_{J_2}\,
\cos[m(\phi_{J_1}-\phi_{J_2}-\pi)] \,
\frac{d\sigma}{dy_{J_1}dy_{J_2}\, d|\vec k_{J_1}| \, d|\vec k_{J_2}|
d\phi_{J_1} d\phi_{J_2}}\;.
\eeq

In particular, taking into account the Jacobian of the transformation from the
variables $\vec k_{J_i}$, $x_{J_i}$ to the variables $|\vec k_{J_i}|$,
$y_{J_i}$, and the $\nu$-dependence of LLA jet vertices, see~(\ref{v1}), we get
\beq{C0}
{\cal C}_n
= \frac{x_{J_1} x_{J_2}}{|\vec k_{J_1}| |\vec k_{J_2}|}
\int_{-\infty}^{+\infty} d\nu \, \left(\frac{x_{J_1} x_{J_2} s}{s_0}
\right)^{\bar \alpha_s(\mu_R)\chi(n,\nu)}
\eeq
\[
\times \alpha_s^2(\mu_R) c_1(n,\nu,|\vec k_{J_1}|, x_{J_1})
c_2(n,\nu,|\vec k_{J_2}|,x_{J_2})\,
\]
\[
\times \left[1
+\alpha_s(\mu_R)\left(\frac{c_1^{(1)}(n,\nu,|\vec k_{J_1}|,
x_{J_1})}{c_1(n,\nu,|\vec k_{J_1}|, x_{J_1})}
+\frac{c_2^{(1)}(n,\nu,|\vec k_{J_2}|, x_{J_2})}{c_2(n,\nu,|\vec k_{J_2}|,
x_{J_2})}\right)
\right.
\]
\[
\left.
+\bar \alpha_s^2(\mu_R) \ln\left(\frac{x_{J_1}x_{J_2} s}{s_0}\right)
\left(\bar \chi(n,\nu)
+ \frac{\beta_0}{8C_A}\chi(n,\nu)
\left(-\chi(n,\nu) + \frac{10}{3} + \ln\frac{\mu_R^4}
{\vec k_{J_1}^2 \vec k_{J_2}^2}\right)\right)\right] \;,
\]
where
\beq{c1}
c_1(n,\nu,|\vec k|,x)=2\sqrt{\frac{C_F}{C_A}}
(\vec k^{\,2})^{i\nu-1/2}\,\left(\frac{C_A}{C_F}f_g(x,\mu_F)
+\sum_{a=q,\bar q}f_q(x,\mu_F)\right) \;,
\eeq
\beq{c2}
c_2(n,\nu,|\vec k|,x)=\biggl[c_1(n,\nu,|\vec k|,x)
\biggr]^* \;,
\eeq
\beq{c11}
c_1^{(1)}(n,\nu,|\vec k|,x)=
\frac{1}{\pi}\sqrt{\frac{C_F}{C_A}}
\left(\vec k^{\,2} \right)^{i\nu-1/2}
\int\limits^1_{x}\frac{d\zeta}{\zeta}
\zeta^{-\bar\alpha_s(\mu_R)\chi(n,\nu)}
\left\{\sum_{a=q,\bar q} f_a \left(\frac{x}{ \zeta}\right)\right.
\eeq
\[
\times \left[\left(P_{qq}(\zeta)+\frac{C_A}{C_F}P_{gq}(\zeta)\right)
\ln\frac{\vec k^{\,2}}{\mu_F^2}-2\zeta^{-2\gamma}\ln R\,
\left\{P_{qq}(\zeta)+P_{gq}(\zeta)\right\}-\frac{\beta_0}{2}
\ln\frac{\vec k^{\,2}}{\mu_R^2}\delta(1-\zeta)\right.
\]
\[
+C_A\delta(1-\zeta)\left(\chi(n,\gamma)\ln\frac{s_0}{\vec k^{\,2}}
+\frac{85}{18}+\frac{\pi^2}{2}+\frac{1}{2}\left(\psi^\prime
\left(1+\gamma+\frac{n}{2}\right)
-\psi^\prime\left(\frac{n}{2}-\gamma\right)-\chi^2(n,\gamma)\right)
\right)
\]
\[
+(1+\zeta^2)\left\{C_A\left(\frac{(1+\zeta^{-2\gamma})\,\chi(n,\gamma)}
{2(1-\zeta)_+}-\zeta^{-2\gamma}\left(\frac{\ln(1-\zeta)}{1-\zeta}\right)_+
\right)+\left(C_F-\frac{C_A}{2}\right)\left[ \frac{\bar \zeta}{\zeta^2}I_2
-\frac{2\ln\zeta}{1-\zeta}\right.\right.
\]
\[
\left.\left.
+2\left(\frac{\ln(1-\zeta)}{1-\zeta}\right)_+ \right]\right\}+
\delta(1-\zeta)\left(C_F\left(3\ln 2-\frac{\pi^2}{3}-\frac{9}{2}\right)
-\frac{5n_f}{9}\right)
\]
\[
\left. +C_A\zeta+C_F\bar \zeta+\frac{1+\bar \zeta^2}{\zeta}
\left(C_A\frac{\bar \zeta}{\zeta}I_1+2C_A\ln\frac{\bar\zeta}{\zeta}
+C_F\zeta^{-2\gamma}(\chi(n,\gamma)-2\ln \bar \zeta)\right)\right]
\]
\[
+f_{g}\left(\frac{x}{ \zeta}\right)\frac{C_A}{C_F}
\]
\[
\times \left[
\left(P_{gg}(\zeta)+2 \,n_f \frac{C_F}{C_A}P_{qg}(\zeta)\right)
\ln\frac{\vec k^{\,2}}{\mu_F^2}-
2\zeta^{-2\gamma}\ln R \left(P_{gg}(\zeta)+2 \,n_f P_{qg}(\zeta)\right)
-\frac{\beta_0}{2}\ln\frac{\vec k^{\,2}}{4\mu_R^2}\delta(1-\zeta)
\right.
\]
\[
+\, C_A\delta(1-\zeta)
\left(
\chi(n,\gamma)\ln\frac{s_0}{\vec k^{\,2}}+\frac{1}{12}+\frac{\pi^2}{6}
+\frac{1}{2}\left(\psi^\prime\left(1+\gamma+\frac{n}{2}\right)
-\psi^\prime\left(\frac{n}{2}-\gamma\right)-\chi^2(n,\gamma)\right)
\right)
\]
\[
%+\,
%\delta(1-\zeta)\, \ln 2 \left(\frac{11 C_A}{3}-\frac{2n_f}{3}\right)
+\, 2 C_A (1-\zeta^{-2\gamma})\left(\left(\frac{1}{\zeta}-2
+\zeta\bar\zeta\right)\ln \bar \zeta + \frac{\ln (1-\zeta)}{1-\zeta}\right)
\]
\[
+ \, C_A\, \left[\frac{1}{\zeta}+\frac{1}{(1- \zeta)_+}-2+\zeta\bar\zeta\right]
\left((1+\zeta^{-2\gamma})\chi(n,\gamma)-2\ln\zeta+\frac{\bar \zeta^2}
{\zeta^2}I_2\right)
\]
\[
\left.\left.
+\, n_f\left[\, 2\zeta\bar \zeta \, \frac{C_F}{C_A} +(\zeta^2+\bar \zeta^2)
\left(\frac{C_F}{C_A}\chi(n,\gamma)+\frac{\bar \zeta}{\zeta}I_3\right)
-\frac{1}{12}\delta(1-\zeta)\right]\right]\right\}\;,
\]
\beq{c21}
c_2^{(1)}(n,\nu,|\vec k|,x)=
\biggl[c_1^{(1)}(n,\nu,|\vec k|,x)\biggr]^*\;.
\eeq
Here $\bar \zeta=1-\zeta$, $\gamma=i\nu-1/2$, $P_{i j}(\zeta)$ are leading
order DGLAP kernels.
For the $I_{1,2,3}$ functions we have the results:
\beq{}
I_2=\frac{\zeta^2}{\bar \zeta^2}\left[
\zeta\left(\frac{{}_2F_1(1,1+\gamma-\frac{n}{2},2+\gamma-\frac{n}{2},\zeta)}
{\frac{n}{2}-\gamma-1}-
\frac{{}_2F_1(1,1+\gamma+\frac{n}{2},2+\gamma+\frac{n}{2},\zeta)}{\frac{n}{2}+
\gamma+1}\right)\right.
\eeq
$$
\left.
+\zeta^{-2\gamma}
\left(\frac{{}_2F_1(1,-\gamma-\frac{n}{2},1-\gamma-\frac{n}{2},\zeta)}
{\frac{n}{2}+\gamma}-
\frac{{}_2F_1(1,-\gamma+\frac{n}{2},1-\gamma+\frac{n}{2},\zeta)}{\frac{n}{2}
-\gamma}\right)
\right.
$$
%$$
%\left.
%+\zeta^{\frac{n}{2}-\gamma }\left(
%\psi\left(\gamma+\frac{n}{2}\right)+\psi\left(1-\gamma+\frac{n}{2}\right)
%-\psi\left(\gamma-\frac{n}{2}\right)-\psi\left(1-\gamma-\frac{n}{2}\right)
%\right)
%\right.
%$$
$$
\left.
+\left(1+\zeta^{-2\gamma}\right)\left(\chi(n,\gamma)-2\ln \bar \zeta \right)
+2\ln{\zeta}\right]\;,
$$

\beq{}
I_1=\frac{\bar \zeta}{2\zeta}I_2+\frac{\zeta}{\bar \zeta}\left[
\ln \zeta+\frac{1-\zeta^{-2\gamma}}{2}\left(\chi(n,\gamma)-2\ln \bar \zeta
\right)\right]\;,
\eeq

\beq{}
I_3=\frac{\bar \zeta}{2\zeta}I_2-\frac{\zeta}{\bar \zeta}\left[
\ln \zeta+\frac{1-\zeta^{-2\gamma}}{2}\left(\chi(n,\gamma)-2\ln \bar \zeta
\right)\right]\;.
\eeq
The factor $\zeta^{-\bar\alpha_s(\mu_R)\chi(n,\nu)}$ appears in~(\ref{c11})
due to extra-contributions attributed to the jet vertices, as discussed after
Eq.~(\ref{factors}). Note that the ${\cal C}_n$ coefficients  do not
depend on the azimuthal angles of the jets, $\phi_{J_1}$ and $\phi_{J_2}$,
\beq{c-n-dep}
{\cal C}_n={\cal C}_n\left(y_{J_1},y_{J_2},\vec k_{J_1},\vec k_{J_2},\mu_R,
\mu_F,s_0\right) \; ,
\eeq
they depend instead on the jet rapidities, the transverse momenta and on
the factorization, renormalization and energy scale parameters.

An alternative way to present the differential cross section, equivalent to
the formula~(\ref{C0}) in the NLA, is the so-called {\it exponentiated} form
(see Ref.~\cite{mesons}),
\beq{C0_exp}
{\cal C}^{\rm exp}_n
= \frac{x_{J_1} x_{J_2}}{|\vec k_{J_1}| |\vec k_{J_2}|}
\int_{-\infty}^{+\infty} d\nu \, \exp\biggl[(Y-Y_0)
\biggl(\bar \alpha_s(\mu_R)\chi(n,\nu)\biggr.\biggr.
\eeq
\[
\left.\left.
+\bar \alpha_s^2(\mu_R)\left(\bar \chi(n,\nu)
+ \frac{\beta_0}{8C_A}\chi(n,\nu)
\left(-\chi(n,\nu) + \frac{10}{3} + \ln\frac{\mu_R^4}
{\vec k_{J_1}^2 \vec k_{J_2}^2}\right)\right)\right)\right]
\]
\[
\times  \alpha_s^2(\mu_R)  c_1(n,\nu,|\vec k_{J_1}|, x_{J_1},\mu_F)
c_2(n,\nu,|\vec k_{J_2}|,x_{J_2},\mu_F)\,
\]
\[
\times\left[1
 +\alpha_s(\mu_R)\left(\frac{c_1^{(1)}(n,\nu,|\vec k_{J_1}|,
x_{J_1},\mu_F)}{c_1(n,\nu,|\vec k_{J_1}|,x_{J_1},\mu_F)}
+\frac{c_2^{(1)}(n,\nu,|\vec k_{J_2}|,x_{J_2},
\mu_F)}{c_2(n,\nu,|\vec k_{J_2}|,x_{J_2},\mu_F)}\right)\right]\;,
\]
where $Y$ is defined in Eq.~(\ref{Y}) and we have introduced the variable
\beq{Y0}
Y_0=\ln{\left({s_0\over |\vec k_{J_1}||\vec k_{J_2}|}\right)} \, .
\eeq

Below we will discuss the differential cross section integrated over the jet
azimuthal angles
\[
{\cal C}_0 = \int d\phi_{J_1}d\phi_{J_2} d\sigma\;,
\]
the coefficients ${\cal C}_n$ and  the moments of the azimuthal
decorrelations, which are defined as
\beq{decorr}
\langle \cos(n \phi) \rangle =\frac{\int d\phi_{J_1}d\phi_{J_2}
\cos[n(\phi_{J_1}-\phi_{J_2}-\pi)] d\sigma}
{\int d\phi_{J_1}d\phi_{J_2}  d\sigma} = \frac{{\cal C}_n}{{\cal C}_0}\;.
\eeq

%C_0 plot
\begin{figure}[b]
\centering
\includegraphics[scale=0.55]{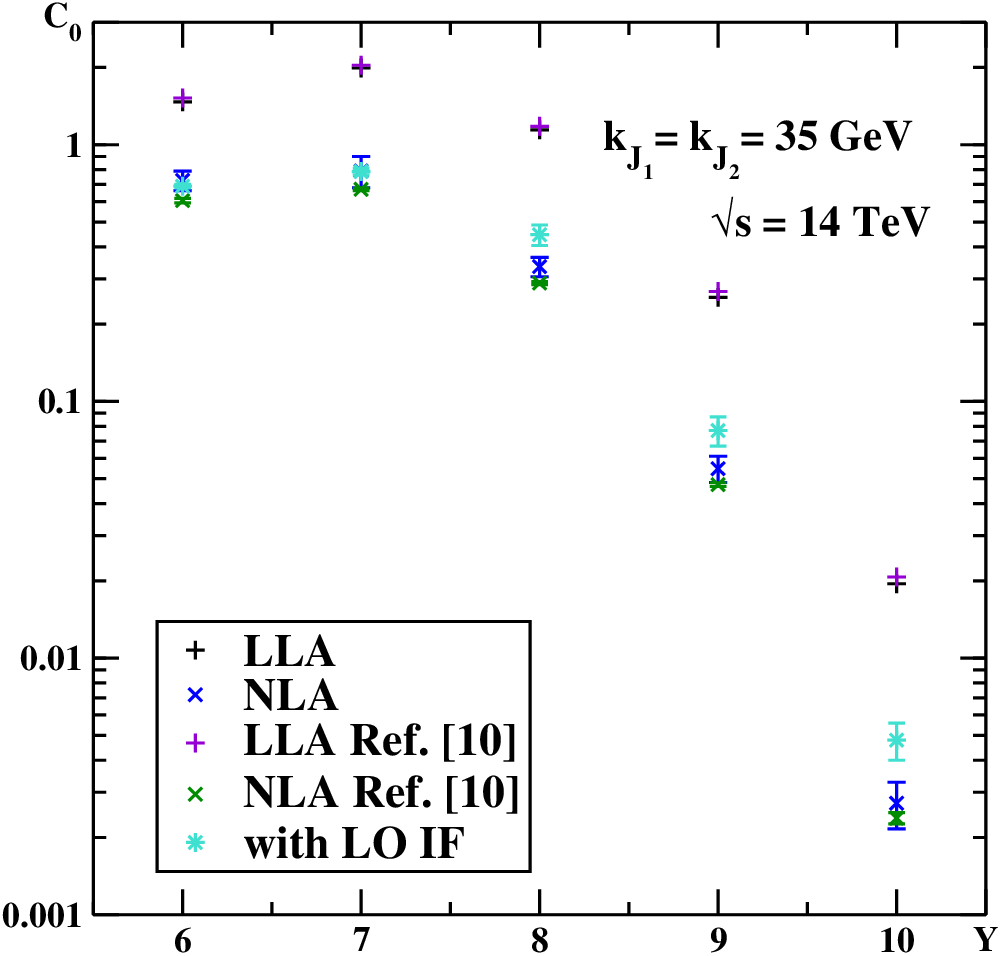}
\caption[]{$Y$ dependence of $C_0$ for $|\vec k_{J_1}|=|\vec k_{J_2}|=35$
GeV at $\sqrt s=14$ TeV.}
\label{fig:C035}
\end{figure}
%C_0 table
\begin{table}[h]
\begin{center}
\begin{tabular}{|r|c|c|c|c||c|c|c|}
\hline
$Y$ & $C_0^{(\rm LLA)}$ & $C_0^{(\rm NLA)}$ & $Y_0$ & $n_R$
    & $C_0^{(\rm NLA/LO \ IF)}$ & $Y_0$ & $n_R$ \\
\hline
6  & 1.468    & 0.726(64)   & 1 & 2 & 0.689(11)   & 1 & 12 \\
7  & 1.990    & 0.79(11)    & 2 & 2 & 0.786(23)   & 1 & 6  \\
8  & 1.142    & 0.335(29)   & 2 & 2 & 0.446(41)   & 1 & 2  \\
9  & 0.2542   & 0.0547(64)  & 3 & 2 & 0.077(10)   & 1 & 2  \\
10 & 0.01947  & 0.00272(56) & 4 & 2 & 0.00479(79) & 1 & 2  \\
\hline
\end{tabular}
\end{center}
\caption{Values of $C_0$ in the LLA, in the NLA and in the NLA with
LO impact factors for $|\vec k_{J_1}|=|\vec k_{J_2}|=35$ GeV at $\sqrt s=14$
TeV, corresponding to the data points in Fig.~\ref{fig:C035}. The
optimal values of $Y_0$ and $n_R=\mu_R/ \sqrt{|\vec k_{J_1}| |\vec k_{J_2}|}$
for $C_0^{(\rm NLA)}$ are given in the fourth and fifth columns, while
those for $C_0^{(\rm NLA/LO \ IF)}$ are given in the last two columns.}
\label{tab:C035}
\end{table}

%%%%%%%%%%%%%%%%%%%%%%%%%%%%%%%%%%%%%%%%%%%%%%%%%%%%%%%%%%%%%%%%%%%%%%%%%%%

%C_0 plot
\begin{figure}[t]
\centering
\includegraphics[scale=0.55]{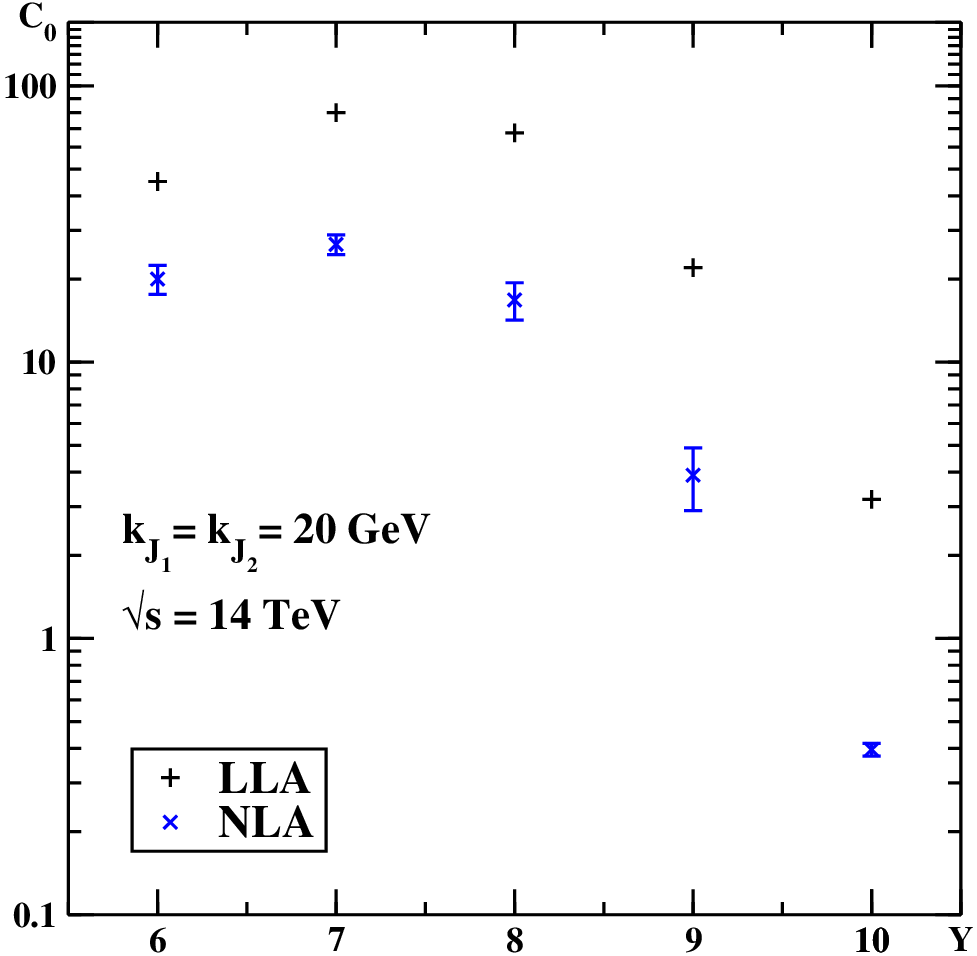}
\caption[]{$Y$ dependence of $C_0$ for $|\vec k_{J_1}|=|\vec k_{J_2}|=20$ GeV
at $\sqrt s=14$ TeV.}
\label{fig:C020}
\end{figure}
%C_0 table
\begin{table}[h]
\begin{center}
\begin{tabular}{|r|c|c|c|c|}
\hline
$Y$ & $C_0^{(\rm LLA)}$ & $C_0^{(\rm NLA)}$ & $Y_0$ & $n_R$\\
\hline
6  & 45.12 & 20.0(24)  & 1 & 1\\
7  & 80.03 & 26.7(22)  & 1 & 1\\
8  & 67.65 & 16.8(26)  & 2 & 1\\
9  & 21.99 & 3.9(10)   & 3 & 1\\
10 & 3.187 & 0.396(21) & 3 & 2\\
\hline
\end{tabular}
\end{center}
\caption{Values of $C_0$ in the LLA and in the NLA for
$|\vec k_{J_1}|=|\vec k_{J_2}|=20$ GeV at $\sqrt s=14$ TeV, corresponding to
the data points in Fig.~\ref{fig:C020}. The last two columns give the optimal
values of $Y_0$ and $n_R=\mu_R/ \sqrt{|\vec k_{J_1}| |\vec k_{J_2}|}$.}
\label{tab:C020}
\end{table}

\section{Results}

In this section we present our results for the dependence on $Y$ of the
functions ${\cal C}_n$. In what follows we take the factorization and
renormalization scales equal to each other, $\mu_F=\mu_R$.
We perform our calculation both in the LLA and in the NLA. In the former case,
the expression for ${\cal C}_n$ reads
\beq{C0_LLA}
{\cal C}^{\rm{LLA}}_n=\frac{d\sigma_{\rm LLA}}{dy_{J_1}dy_{J_2}d|\vec k_{J_1}|
\, d|\vec k_{J_2}|}
= \frac{x_{J_1} x_{J_2}}{|\vec k_{J_1}| |\vec k_{J_2}|}
\int_{-\infty}^{+\infty} d\nu \, \exp\biggl[(Y-Y_0)\bar \alpha_s(\mu_R)
\chi(n,\nu)\biggr]
\eeq
\[
\times \alpha_s^2(\mu_R) c_1(n,\nu,|\vec k_{J_1}|,x_{J_1})
c_2(n,\nu,|\vec k_{J_2}|,x_{J_2})\;.
\]
For our NLA analysis we use the exponentiated representation given in
Eq.~(\ref{C0_exp}).

For the center-of-mass energy $\sqrt{s}$ we take the LHC design value
14 TeV. We fix the jet cone size at the value $R=0.5$, in order to compare
our predictions with the forthcoming LHC data. We study Mueller-Navelet
jets with symmetric values  of the transverse momenta, in
particular, consider the choices: $|\vec k_{J_1}|=|\vec k_{J_2}|=35$ GeV
and $|\vec k_{J_1}|=|\vec k_{J_2}|=20$ GeV. 

Moreover, to make possible the comparison with the experiments at present
LHC energy, we perform calculations for $\sqrt{s}=7$ TeV, where we consider
the $|\vec k_{J_1}|=|\vec k_{J_2}|=35$ GeV case.

Following a quite recent CMS study~\cite{CMS}, we restrict the rapidities of
the Mueller-Navelet jets to the region $3 \leq |y_J | \leq 5$. We will present
our results for ${\cal C}_0$, {\it i.e.} the differential cross section
integrated over the jet azimuthal angles, the coefficients
${\cal C}_n$ for $n\neq 0$, and $\langle \cos(n \phi) \rangle$ versus the
relative rapidity, $Y = y_{J_1}  - y_{J_2}$. For our choice of forward jets
rapidities, $Y$ takes values between 6 and 10. Our approach is similar to the
one used in~\cite{Colferai:2010wu}, we introduce rapidity bins with step
$\Delta y_J$ equal to 0.5, so the considered values for jet rapidities and
rapidity difference are
\begin{eqnarray*}
\{(y_{J_1})_i\}&=&\{3.0,3.5,4.0,4.5,5.0\} \, \\
\{(y_{J_2})_i\}&=&\{-3.0,-3.5,-4.0,-4.5,-5.0\}
\end{eqnarray*}
and $\{Y_i\}=\{6.0,6.5,7.0,7.5,8.0,8.5,9.0,9.5,10.0\}$.
Then we evaluate the following sum
$$
C_n(Y_i)=\sum 0.5\, {\cal C}_n\left( (y_{J_1})_j,(y_{J_1})_j-Y_i\right)
$$
where the sum runs over the possible values of $(y_{J_1})_j$ for a given $Y_i$.

%%%%%%%%%%%%%%%%%%%%%%%%%% 35 GeV - \sqrt s= 7 TeV%%%%%%%%%%%%%%%%%%%%%%%%%%%%%

%C_0 plot
\begin{figure}[b]
\centering
\includegraphics[scale=0.55]{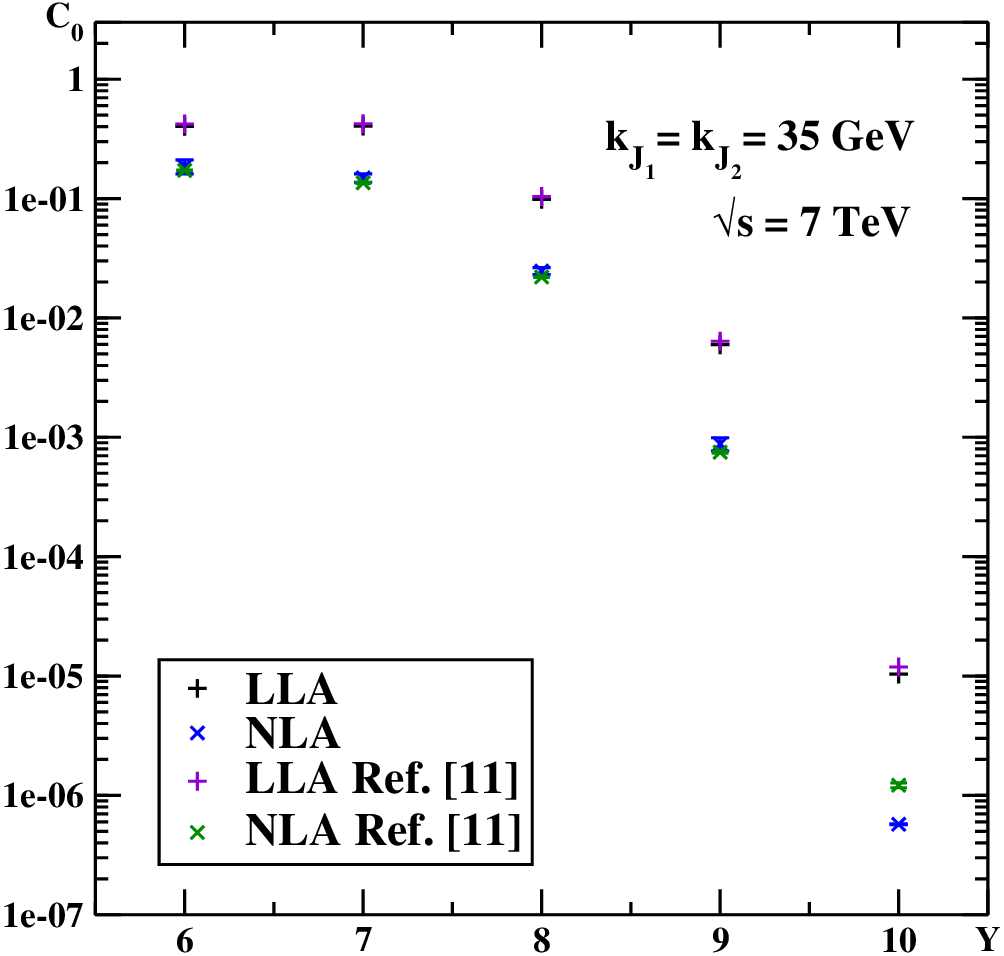}
\caption[]{$Y$ dependence of $C_0$ for $|\vec k_{J_1}|=|\vec k_{J_2}|=35$ GeV
at $\sqrt{s}=7$ TeV.}
\label{fig:C035s7}
\end{figure}

%C_0 table
\begin{table}[h]
\begin{center}
\begin{tabular}{|r|c|c|c|c|}
\hline
$Y$ & $C_0^{(\rm LLA)}$ & $C_0^{(\rm NLA)}$ & $Y_0$ & $n_R$\\
\hline
6  & 0.403            & 0.186(25)          & 2 & 2\\
7  & 0.405            & 0.149(12)          & 2 & 3\\
8  & 0.0984           & 0.0247(17)         & 3 & 3\\
9  & 0.005974         & 0.00088(11)        & 4 & 3\\
10 & 103.62 $10^{-7}$ & 5.73(51) $10^{-7}$ & 5.5 & 7\\
\hline
\end{tabular}
\end{center}
\caption{Values of $C_0$ in the LLA and in the NLA for $|\vec k_{J_1}|
=|\vec k_{J_2}|=35$ GeV at $\sqrt{s}=7$ TeV, corresponding to the data points
in Fig.~\ref{fig:C035s7}. The last two columns give the optimal values of
$Y_0$ and $n_R=\mu_R/ \sqrt{|\vec k_{J_1}| |\vec k_{J_2}|}$.}
\label{tab:C035s7}
\end{table}

In our analysis we use the PDF set MSTW2008nnlo~\cite{pdf} and the two-loop
running coupling with $\alpha_s(M_Z)=0.11707.$

%%%%%%%%%%%%%%%%%%%%%%%%%%%%%%%%%%%%%%%%%%%%%%%%%%%%%%%%%%%%%%%%%%%%%%%%%%%

%C_1 + C_2 plot (35 Gev)
\begin{figure}[t]
\centering
\includegraphics[scale=0.45]{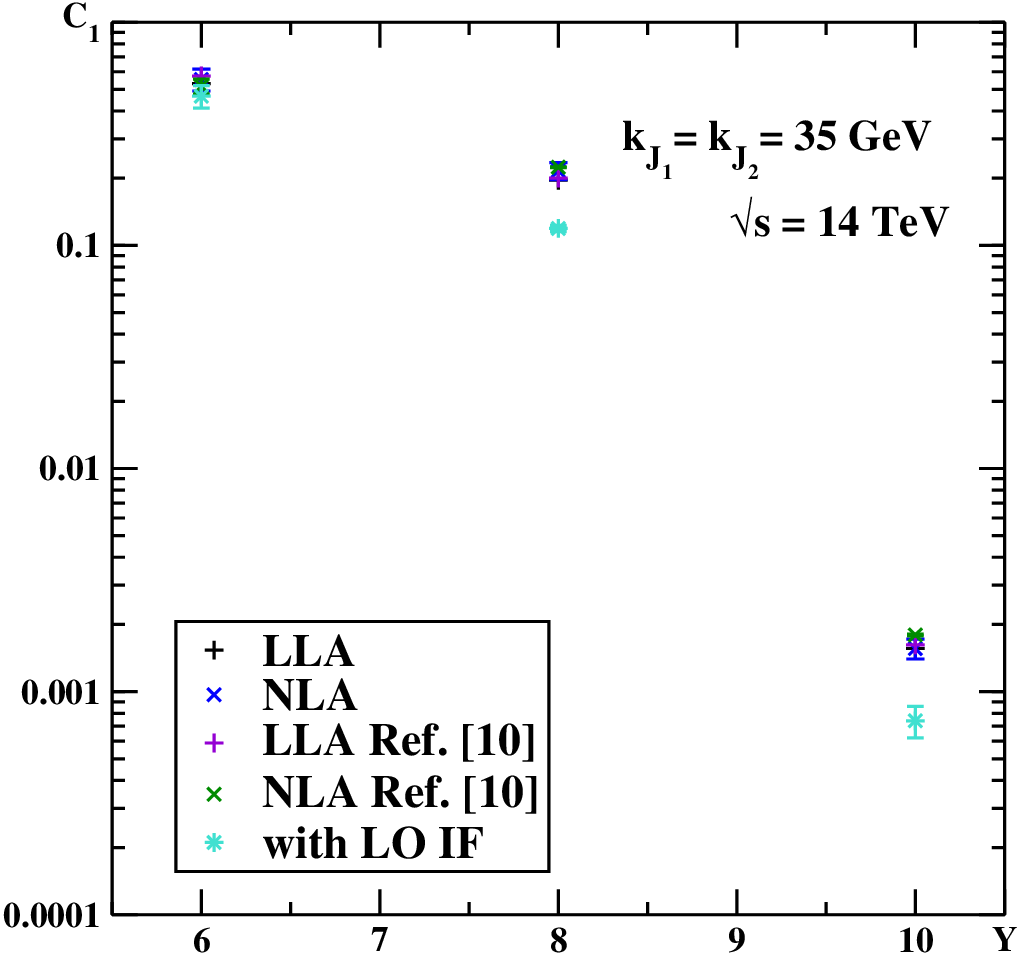}
\hspace{0.4cm}
\includegraphics[scale=0.45]{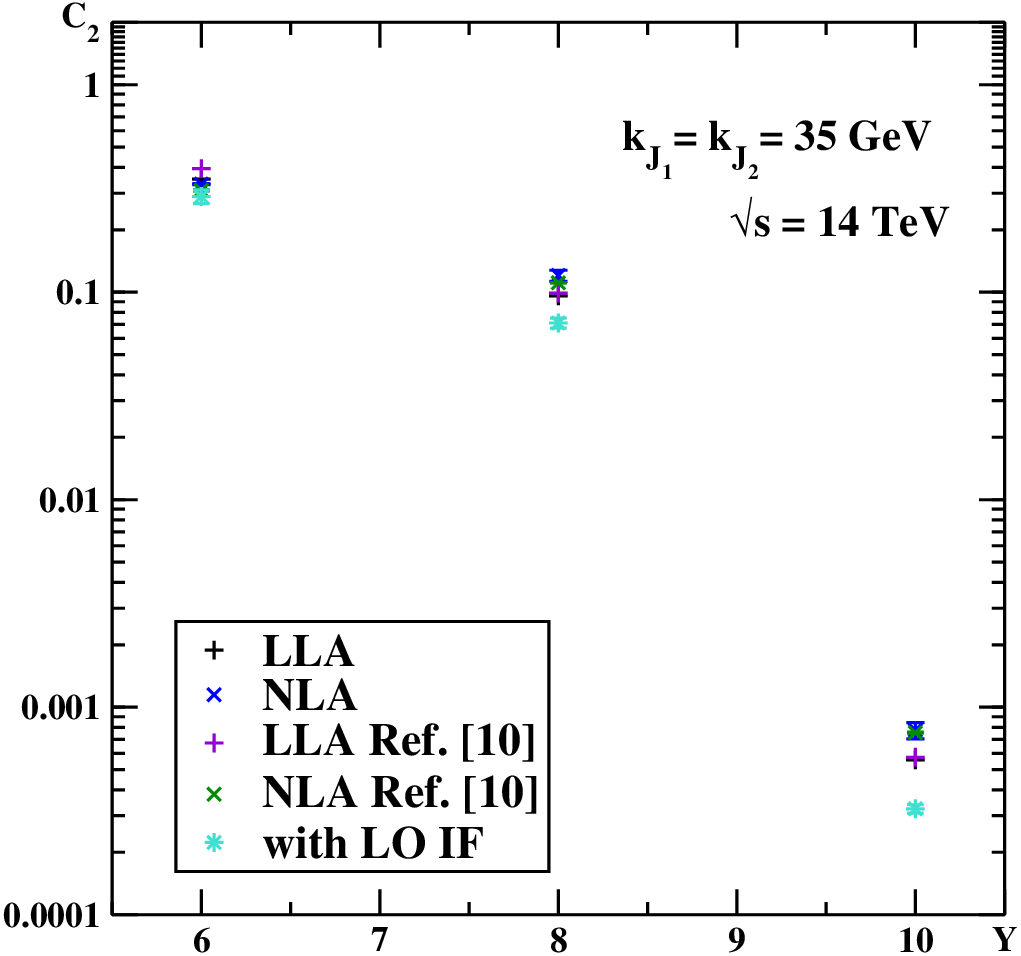}
\caption[]{$Y$ dependence of $C_1$ (left) and $C_2$ (right) for
$|\vec k_{J_1}|=|\vec k_{J_2}|=35$ GeV at $\sqrt s=14$ TeV.}
\label{fig:C1C235}
\end{figure}
%C_1 + C_2 table
\begin{table}[h]
\begin{center}
\begin{tabular}{|r|c|c|c|c||c|c|c|c|}
\hline
$Y$ & $C_1^{(\rm LLA)}$ & $C_1^{(\rm NLA)}$ & $Y_0$ & $n_R$
    & $C_2^{(\rm LLA)}$ & $C_2^{(\rm NLA)}$ & $Y_0$ & $n_R$\\
\hline
6  & 0.531   & 0.554(62)   & 1 & 2 & 0.351    & 0.3320(18)   & 0 & 1.5 \\
8  & 0.196   & 0.216(19)   & 2 & 2 & 0.0961   & 0.1203(74)   & 2 & 2.5 \\
10 & 0.00156 & 0.00156(16) & 3 & 2 & 0.000558 & 0.000774(69) & 4 & 4   \\
\hline
\end{tabular}
\end{center}
\caption{Values of $C_1$ and $C_2$ in the LLA and in the NLA for
$|\vec k_{J_1}|=|\vec k_{J_2}|=35$ GeV at $\sqrt s=14$ TeV, corresponding to
the data points in Fig.~\ref{fig:C1C235}. The optimal values of $Y_0$ and
$n_R=\mu_R/ \sqrt{|\vec k_{J_1}| |\vec k_{J_2}|}$ for $C_1^{(\rm NLA)}$
are given in the fourth and fifth columns, while those for $C_2^{(\rm NLA)}$ are
given in the last two columns.}
\label{tab:C1C235}
\end{table}

%%%%%%%%%%%%%%%%%%%%%%%%%%%%%%%%%%%%%%%%%%%%%%%%%%%%%%%%%%%%%%%%%%%%%%%%%%%

%C_1 + C_2 plot (20 Gev)
\begin{figure}[b]
\centering
\includegraphics[scale=0.45]{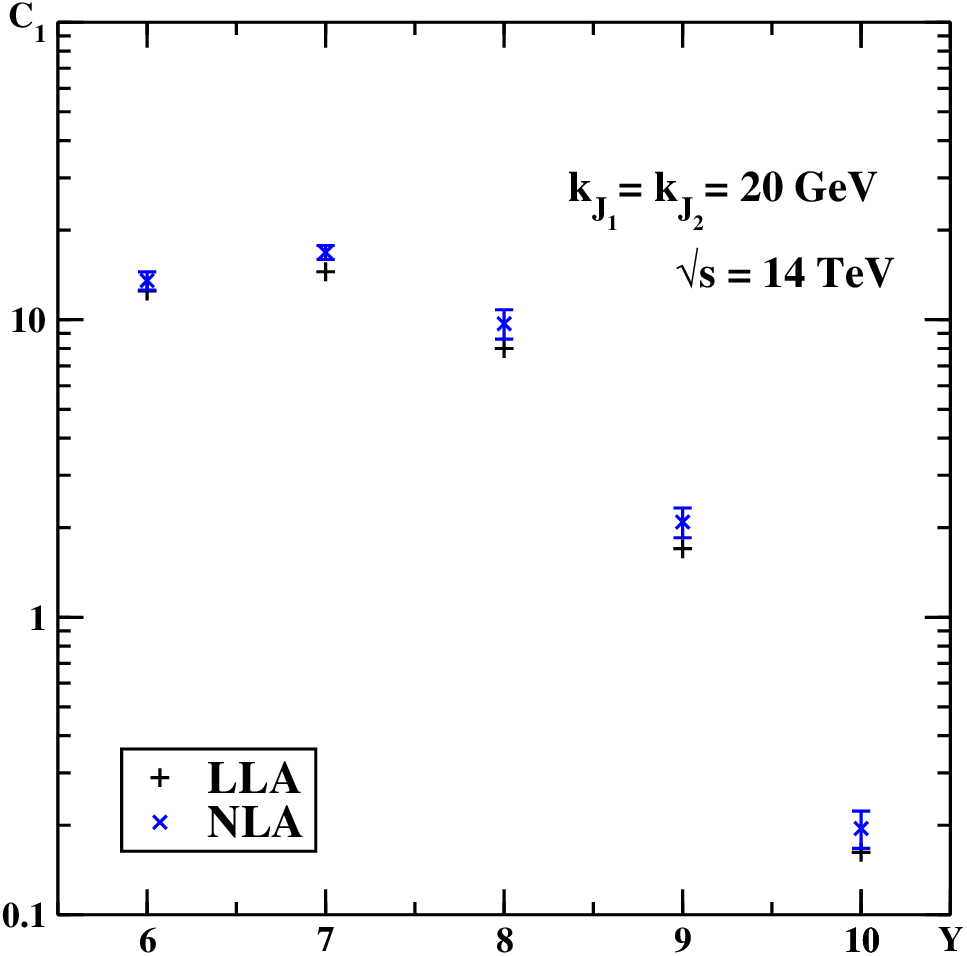}
\hspace{0.55cm}
\includegraphics[scale=0.45]{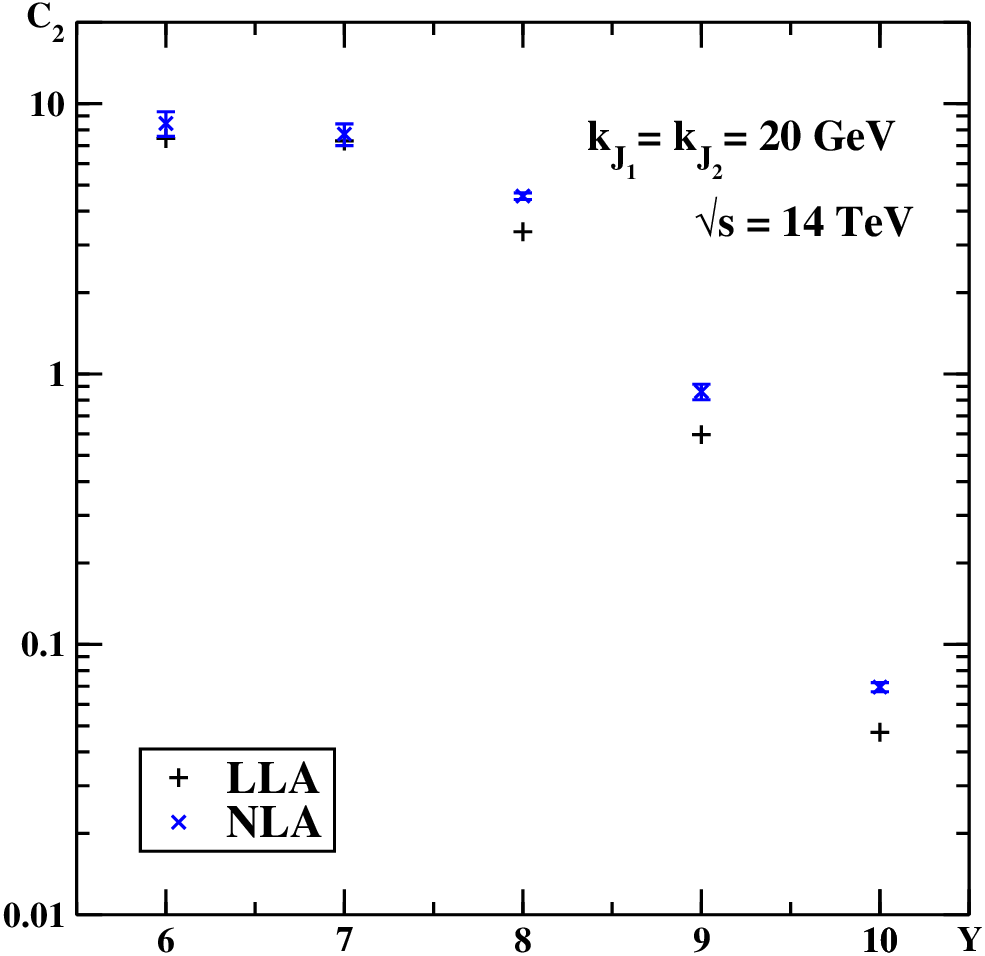}
\caption[]{$Y$ dependence of $C_1$ (left) and $C_2$ (right) for
$|\vec k_{J_1}|=|\vec k_{J_2}|=20$ GeV at $\sqrt s=14$ TeV.}
\label{fig:C1C220}
\end{figure}
%C_1 + C_2 table
\begin{table}[h]
\begin{center}
\begin{tabular}{|r|c|c|c|c||c|c|c|c|}
\hline
$Y$ & $C_1^{(\rm LLA)}$ & $C_1^{(\rm NLA)}$ & $Y_0$ & $n_R$
    & $C_2^{(\rm LLA)}$ & $C_2^{(\rm NLA)}$ & $Y_0$ & $n_R$\\
\hline
6  & 12.49  & 13.54(95)  & 0   & 1 & 7.43   & 8.45(88)   & 1   & 3 \\
7  & 14.50  & 16.85(91)  & 0.5 & 1 & 7.28   & 7.71(71)   & 0.5 & 1 \\
8  &  8.0   &  9.7(11)   & 1   & 1 & 3.36   & 4.55(13)   & 2.5 & 4 \\
9  &  1.70  &  2.09(24)  & 1   & 1 & 0.597  & 0.860(57)  & 3   & 5 \\
10 &  0.162 &  0.195(28) & 1   & 1 & 0.0473 & 0.0695(27) & 3.5 & 3 \\
\hline
\end{tabular}
\end{center}
\caption{Values of $C_1$ and $C_2$ in the LLA and in the NLA for
$|\vec k_{J_1}|=|\vec k_{J_2}|=20$ GeV at $\sqrt s=14$ TeV, corresponding to
the data points in Fig.~\ref{fig:C1C220}. The optimal values of $Y_0$ and
$n_R=\mu_R/ \sqrt{|\vec k_{J_1}| |\vec k_{J_2}|}$ for $C_1^{(\rm NLA)}$
are given in the fourth and fifth columns, while those for $C_2^{(\rm NLA)}$ are
given in the last two columns.}
\label{tab:C1C220}
\end{table}

Our predictions depend on the values of energy and renormalization scales,
$s_0$ and $\mu_R$. For the analysis in the LLA we fixed the values of these
scales, $\mu_R$ and $s_0$, as suggested by the kinematics of the process,
{\it i.e.} $\mu_R^2=s_0=|\vec k_{J_1}||\vec k_{J_2}|$.

In general LLA results depend very strongly on $s_0$ and $\mu_R$, and one
really needs to proceed to the NLA analysis in order to reduce this scale
dependence and to have some reliable predictions for observables.
One should stress that the dependence of the correlations $C_n$ on the scales
$\mu_R$ and $s_0$ cancels with NLA accuracy; nevertheless in both
representations~(\ref{C0}) and~(\ref{C0_exp}) there unavoidably exist
contributions subleading to NLA, depending on $\mu_R$ and $s_0$, whose
numerical impact is important for the considered kinematics, therefore we need
some prescription for the choice of these scales.

Following Ref.~\cite{mesons}, we use here an adaptation of the
{\it principle of minimal sensitivity} (PMS)~\cite{Stevenson}, which consists
in taking as optimal choices for $\mu_R$ and $s_0$ those values for which
the physical observable under examination exhibits the minimal sensitivity to
changes of both of these scales. The motivation of this procedure
is that the complete resummation of the perturbative series would not depend on
the scales $\mu_R$ and $s_0$, so the optimization method is supposed to
mimic the effect of the most relevant unknown subleading terms.

%%%%%%%%%%%%%%%%%%%%%%%%%% 35 GeV %%%%%%%%%%%%%%%%%%%%%%%%%%%%%%%%%%%%%%%%%

%C_1/C_0 plot
\begin{figure}[t]
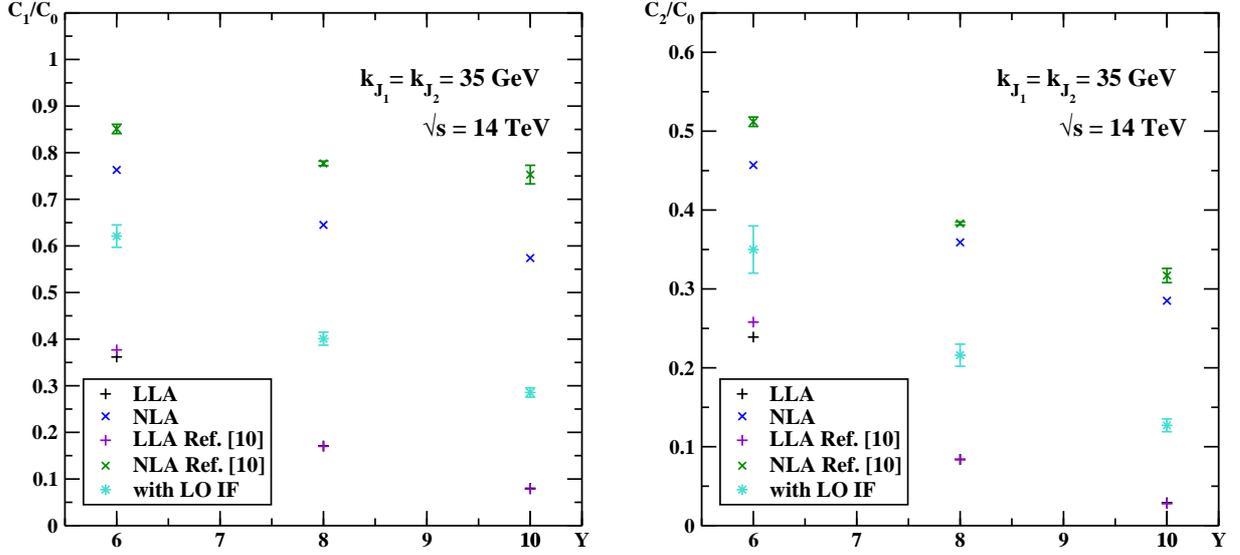

\centering
\includegraphics[scale=0.45]{C1C035.eps}
\hspace{0.4cm}
%C_2/C_0 plot
\includegraphics[scale=0.45]{C2C035.eps}
\caption[]{$Y$ dependence of $C_1/C_0$ (left)
and $C_2/C_0$ (right) for $|\vec k_{J_1}|=|\vec k_{J_2}|=35$ GeV at
$\sqrt s=14$ TeV.   Please note that LLA data at $Y=8$ and 10 from the
present work and from Ref.~\cite{Colferai:2010wu} are barely distinguishable
in these plots.}
\label{fig:C1C035}
\end{figure}
%C_1/C_0 table
\begin{table}[h]
\begin{center}
\begin{tabular}{|r|c|c||c|c|c|}
\hline
$Y$ & $C_1^{(\rm LLA)}/C_0^{(\rm LLA)}$
    & $C_1^{(\rm NLA)}/C_0^{(\rm NLA)}$
    & $C_1^{(\rm NLA / LO \ IF)}/C_0^{(\rm NLA / LO \ IF)}$
    & $Y_0$ & $n_R$ \\
\hline
6  & 0.3618 & 0.763 & 0.621(24)  & 2 & 10 \\
8  & 0.171  & 0.645 & 0.401(14)  & 2 & 5  \\
10 & 0.080  & 0.574 & 0.2854(98) & 3 & 5  \\
\hline
\end{tabular}
\end{center}
\caption{Values of $C_1/C_0=\langle\cos\phi\rangle$ in the LLA, in the NLA
and in the NLA with LO impact factors for $|\vec k_{J_1}|=|\vec k_{J_2}|=35$
GeV at $\sqrt s=14$ TeV, corresponding to the data points in
Fig.~\ref{fig:C1C035}(left). The optimal values of $Y_0$ and
$n_R=\mu_R/ \sqrt{|\vec k_{J_1}| |\vec k_{J_2}|}$ for
$C_1^{(\rm NLA / LO \ IF)}/C_0^{(\rm NLA / LO \ IF)}$ are
given in the last two columns.}
\label{tab:C1C035}
\end{table}
%C_2/C_0 table
\begin{table}[h]
\begin{center}
\begin{tabular}{|r|c|c||c|c|c|}
\hline
$Y$ & $C_2^{(\rm LLA)}/C_0^{(\rm LLA)}$
    & $C_2^{(\rm NLA)}/C_0^{(\rm NLA)}$
    & $C_2^{(\rm NLA / LO \ IF)}/C_0^{(\rm NLA / LO \ IF)}$
    & $Y_0$ & $n_R$ \\
\hline
6  & 0.239 & 0.457 & 0.350(30)  & 1 & 3  \\
8  & 0.084 & 0.359 & 0.216(14)  & 2 & 7  \\
10 & 0.029 & 0.285 & 0.1271(81) & 3 & 12 \\
\hline
\end{tabular}
\end{center}
\caption{Values of $C_2/C_0=\langle\cos\left(2\phi\right)\rangle$ in the LLA, in the NLA
and in the NLA with LO impact factors for $|\vec k_{J_1}|=|\vec k_{J_2}|=35$
GeV at $\sqrt s=14$ TeV, corresponding to the data points in
Fig.~\ref{fig:C1C035}(right). The optimal values of $Y_0$ and
$n_R=\mu_R/ \sqrt{|\vec k_{J_1}| |\vec k_{J_2}|}$ for
$C_2^{(\rm NLA / LO \ IF)}/C_0^{(\rm NLA / LO \ IF)}$ are
given in the last two columns.}
\label{tab:C2C035}
\end{table}

In our search for optimal values, we took integer values for $Y_0$ in the
range 0 -- 6  and values for $\mu_R$ given as integer multiples of
$\sqrt{|\vec k_{J_1}| |\vec k_{J_2}|}$,
\beq{}
\mu_R=n_R \sqrt{|\vec k_{J_1}| |\vec k_{J_2}|}\,,
\eeq
taking the integer $n_R$ in the range  1 -- 7~\footnote{Except than in a few 
cases (see Tables~\ref{tab:C035}, \ref{tab:C1C035} and~\ref{tab:C2C035})
always related with the approximate expressions combining LO impact factors 
and NLA Green's function, which we considered only for comparative
purposes (see the text below).}. The systematic uncertainty of
the optimization procedure in the determination of observables, which will be
discussed below, originates from the resolution of the grid in the
$n_R$ -- $Y_0$ plane. This uncertainty has been estimated as the standard
deviation of the optimal value from the determinations in the nearest
neighbors of the grid.  The error bars around the NLA data points presented
in the figures below represent this uncertainty.   We did not evaluate
the impact on our predictions of the PDF uncertainties, since we expect it
to be of the same size as determined in Ref.~\cite{Colferai:2010wu}, where
the same PDF set adopted here is used.

%%%%%%%%%%%%%%%%%%%%%%%%%% 20 GeV %%%%%%%%%%%%%%%%%%%%%%%%%%%%%%%%%%%%%%%%%

%C_1/C_0 plot
\begin{figure}[b]
\centering
\includegraphics[scale=0.45]{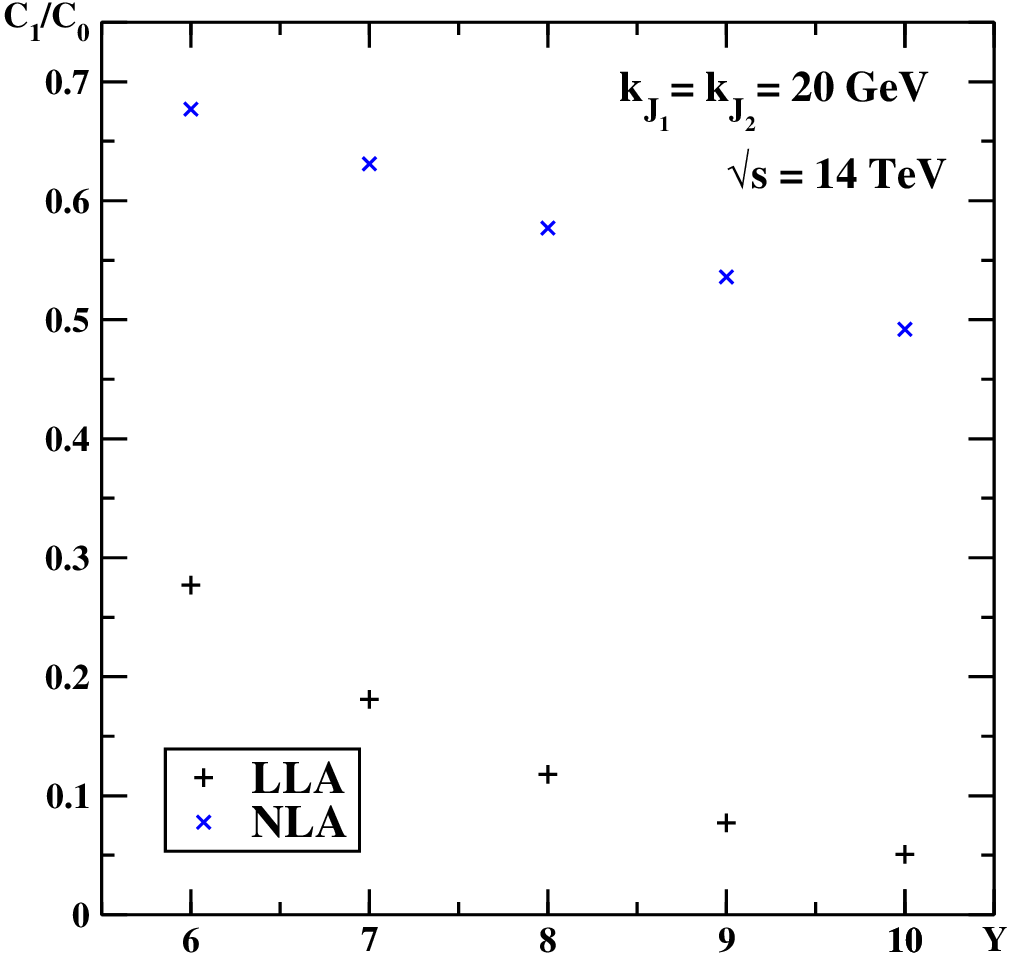}
\hspace{0.4cm}
%C_2/C_0 plot
\includegraphics[scale=0.45]{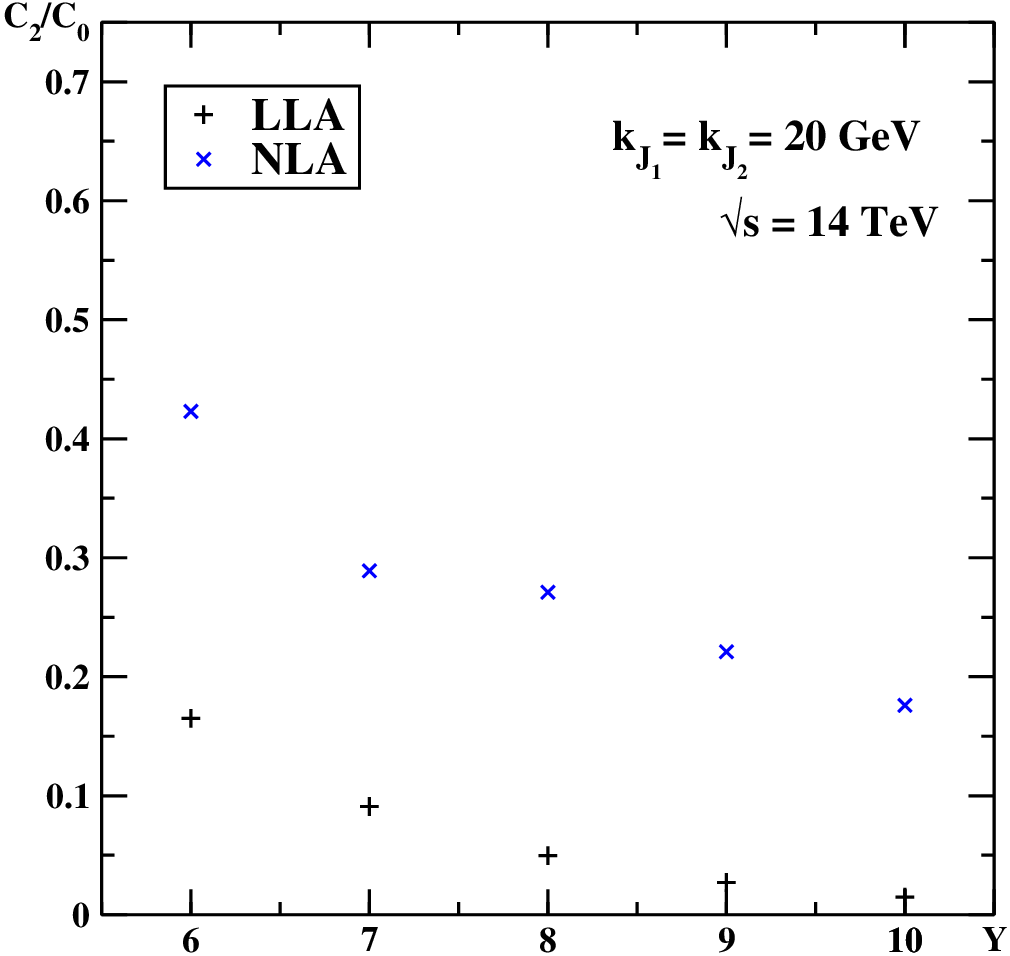}
\caption[]{$Y$ dependence of $C_1/C_0$ (left) and $C_2/C_0$ (right) for
$|\vec k_{J_1}|=|\vec k_{J_2}|=20$ GeV at $\sqrt s=14$ TeV.}
\label{fig:C1C020}
\end{figure}

%C_1/C_0 + C_2/C_0 table
\begin{table}[h]
\begin{center}
\begin{tabular}{|r|c|c||c|c|}
\hline
$Y$ & $C_1^{(\rm LLA)}/C_0^{(\rm LLA)}$
    & $C_1^{(\rm NLA)}/C_0^{(\rm NLA)}$
    & $C_2^{(\rm LLA)}/C_0^{(\rm LLA)}$
    & $C_2^{(\rm NLA)}/C_0^{(\rm NLA)}$ \\
\hline
6  & 0.277  & 0.677 & 0.165 & 0.423 \\
7  & 0.181  & 0.631 & 0.091 & 0.289 \\
8  & 0.118  & 0.577 & 0.050 & 0.271 \\
9  & 0.077  & 0.536 & 0.027 & 0.221 \\
10 & 0.051  & 0.492 & 0.015 & 0.176 \\
\hline
\end{tabular}
\end{center}
\caption{Values of $C_1/C_0=\langle\cos\phi\rangle$ and
$C_2/C_0=\langle\cos\left(2\phi\right)\rangle$ in the LLA and in the NLA
for $|\vec k_{J_1}|=|\vec k_{J_2}|=20$ GeV at $\sqrt s=14$ TeV, corresponding
to the data points in Fig.~\ref{fig:C1C020}.}
\label{tab:C1C020}
\end{table}

Let us start with the cross section integrated over the jet azimuthal angles,
$C_0$. We found that for this observable a stationary point in the
$n_R$ -- $Y_0$ plane could always be singled out, typically a local maximum.
For $\sqrt{s}=14$ TeV our results, in
$\left[\frac{\rm{nb}}{\rm{GeV}^2}\right]$ units, are presented in
Figs.~\ref{fig:C035}--\ref{fig:C020} and in
Tables~\ref{tab:C035}--\ref{tab:C020}; results for $\sqrt{s}=7$ TeV are
given in Fig.~\ref{fig:C035s7} and Table~\ref{tab:C035s7}. The optimal values
of $Y_0$ and $n_R=\mu_R/\sqrt{|\vec k_{J_1}||\vec k_{J_2}|}$ are also
reported in the tables.
As in previous works~\cite{mesons}, the optimal values of the energy scales
turn to be a bit far from the kinematic scale. On the other hand, the uncertainty
related of our optimization procedure, described above, turns out to be small,
therefore our NLA results for the cross section integrated over jet azimuthal
angles, presented in Figs.~\ref{fig:C035}--\ref{fig:C035s7} have relatively
small ``error bars''.

Similar considerations can be done for the observables $C_1$ and $C_2$
at $\sqrt{s}=14$ TeV, to which we refer in
Figs.~\ref{fig:C1C235}--\ref{fig:C1C220}
and in Tables~\ref{tab:C1C235}--\ref{tab:C1C220}. 

The other issue we addressed is the analysis of the
observables ${C}_1/{C}_0$, ${C}_2/{C}_0$, which encode the first two
non-trivial angular decorrelations: $\langle \cos\phi\rangle$ and
$\langle \cos( 2\phi) \rangle$.
For $|\vec k_{J_1}|=|\vec k_{J_2}|=35$ GeV at $\sqrt s=14$ TeV our results are
presented in Fig.~\ref{fig:C1C035} and Tables~\ref{tab:C1C035}
and~\ref{tab:C2C035}.
For the smaller values of jet transverse momenta, $|\vec k_{J_1}|=
|\vec k_{J_2}|=20$ GeV, they are given in Fig.~\ref{fig:C1C020} and
Table~\ref{tab:C1C020}.

In this case, however, we were not able to find clear regions of
stability in the $n_R$ -- $Y_0$ plane, therefore the results we present are
obtained indirectly by using the optimal results for the observables
$C_0$, $C_1$ and $C_2$. 

Even if in general the energy dependence of the cross section and the azimuthal
decorrelations is driven mainly by the kernel, in the considered kinematics
the contribution of the NLO corrections to impact factors happened to be
important.  In order to show this,
in the analysis at $|\vec k_{J_1}|=|\vec k_{J_2}|=35$ GeV
and $\sqrt s=14$ TeV we calculated the coefficient $C_0$ and the
correlations $C_1/C_0$ and $C_2/C_0$ using the NLA BFKL kernel together with
the LO impact factors  (properly modified with the inclusion of the NLO
terms which guarantee the cancellation of the $\mu_R$- and $s_0$-dependence
in the NLA expressions for the coefficients $C_n$, see
Ref.~\cite{Caporale:2008is}). We find that the NLO corrections
to impact factors are relevant especially at large values of Y, as
shown in Figs.~\ref{fig:C035}, \ref{fig:C1C235} and~\ref{fig:C1C035}.

In the following Section we discuss the results presented here, make some
comparison with Refs.~\cite{Colferai:2010wu,Ducloue:2012bm} and draw our
conclusions.

\section{Discussion}

In this paper we considered in NLA BFKL approach the Mueller-Navelet jet
production in proton-proton collisions, using the results for NLA jet
vertices obtained recently in the ``small-cone'' approximation.
Having a simple analytic result for the jet vertices, projected on the
eigenfunction of LLA BFKL equation ($(\nu,n)$-representation), one can
implement them easily in the calculation of the Mueller-Navelet jet cross
section. All necessary formulae are presented in Section~2.

We confirm the observation found earlier in the works devoted to forward
electroproduction of a pair of vector mesons~\cite{mesons}, and to
Mueller-Navelet jets production~\cite{Colferai:2010wu}, that NLA corrections
to the impact factors (jet vertices) are very important and can not be ignored
in a consistent NLA BFKL analysis.

Our numerical results, presented in Section~3, depend on the energy and
renormalization scales, $s_0$ and $\mu_R$. The dependence of the coefficients
${C}_n$ on these scales cancels with NLA accuracy after the inclusion of NLA
corrections to jet vertices. Nevertheless, due to next-to-NLA contributions
depending on $\mu_R$ and $s_0$, the observables we calculated are sensitive
to the choice of these scales.
Here, following Ref.~\cite{mesons}, we used an optimization procedure, based
on the {\it principle of minimal sensitivity}~\cite{Stevenson}, which consists
in taking as optimal choices for $\mu_R$ and $s_0$ those values for which
the physical observable exhibits the minimal sensitivity to changes of both
these scales.

The small-cone approximation, which we adopted here, is expected to be an
adequate tool. Indeed, it is known that in the general case the dependence of
the cross section on the jet cone parameter has, in the limit $R\to 0$, the
form $d\sigma\sim A \ln R+B+{\cal O}(R^2)$ (see, for
instance,~\cite{Furman:1981kf} and Appendix C there). Indeed, in SCA the
coefficients $A$ and $B$ are evaluated exactly. The neglected pieces
${\cal O}(R^2)$ for typical $R$ values are presumably less important than the
other uncertainties of our NLA BFKL calculation, in particular those
related with the choice of the scales $\mu_R$ and $s_0$, which mimic in our
method the effect of the most relevant unknown next-to-NLA BFKL terms.

To support this statement it seems natural to make a comparison of our results
obtained in SCA with the numerics presented
in~\cite{Colferai:2010wu,Ducloue:2012bm}, where the jet cone size was treated
exactly. We would make such comparison for
$|\vec k_{J_1}|=|\vec k_{J_2}|=35$ GeV and  $\sqrt {s}=7$ and 14 TeV.
 For this purpose we present in Table~\ref{tab:comp_Colferai} our NLA
results obtained with the above-discussed optimal scales setting
(presented in the third, fifth and seventh columns) and compare them with
those for $C_0$, $\langle \cos\phi \rangle$ and $\langle \cos (2\phi) \rangle$
taken from Tables~1,~5 and~9 of Ref.~\cite{Colferai:2010wu} and reported
in the second, fourth and sixth columns of our Table~\ref{tab:comp_Colferai}).
Moreover, in Table~\ref{tab:kinematic} we show our NLA results in the case
when kinematic values of the scales were used,
$\mu_R^2=s_0=|\vec k_{J_1}||\vec k_{J_2}|$.

Let us discuss the numbers presented in Tables~\ref{tab:comp_Colferai}
and~\ref{tab:kinematic}. Firstly one needs to say that NLA results obtained
with our formulae at kinematic scale setting,
$\mu_R^2=s_0=|\vec k_{J_1}||\vec k_{J_2}|$, can not be regarded as acceptable
predictions  for high values of $Y$.  In particular  we obtained even a
negative value for the integrated cross section $C_0$ in the case of the largest
rapidity difference, $Y=10$.
This is related to the fact that NLO corrections to the jet vertices are negative
and very large in
absolute value when the kinematic scale setting is used. A similar observation
was done in Refs.~\cite{mesons}, where the electroproduction of a pair of
vector mesons was considered.
This is an indication of the fact that  we actually need the PMS
procedure in order to make reliable predictions.
For $Y<9$ we can see a quite good agreement for $C_0$ with the results of
Ref.~\cite{Colferai:2010wu}, with a discrepancy rising with $Y$. As regards the
moments of azimuthal decorrelation, we can see from Table~\ref{tab:kinematic} that
for the kinematic scales our results for $Y=6$ agree with
Ref.~\cite{Colferai:2010wu}, and also here the discrepancy rises with $Y$.
For the highest value of the rapidity separation, being $C_0$ negative at the
kinematic scales, it makes no sense to report them in the table. 

For the integrated cross section, $C_0$,\footnote{Our LLA results coincide
with those of~\cite{Colferai:2010wu} with high accuracy.  We note, in passing,
that the results quoted in Tables~15, 19, 35, 38, 53 and 56 of
Ref.~\cite{Colferai:2010wu}, giving the coefficients $C_1$ and $C_2$
for several values of the jet kinematics, should be multiplied by a factor
two to correctly reproduce the values of the ratios $C_1/C_0$ and $C_2/C_0$
quoted in other tables of that paper. We stress that this normalization
problem does not affect any of the comparisons presented in this work
between our results and those of Ref.~\cite{Colferai:2010wu}.}
the results presented in the second and third columns of
Table~\ref{tab:comp_Colferai}
 show the same trend, with a numerical discrepancy of about $15\%$. This is due
to the fact that in our approach, with higher values of the scales determined by the
PMS optimization, we effectively take into account part of the subleading
contributions, which are expected to be positive.

Let us discuss now
the observable $\langle \cos\phi \rangle$. Our predictions for it in the case
of $|\vec k_{J_1}|=|\vec k_{J_2}|=35$ GeV and $\sqrt {s}=14$ TeV are shown in
Fig.~\ref{fig:C1C035}(left). In Table~\ref{tab:comp_Colferai} we compare them
(fourth and fifth columns) with those obtained in
Ref.~\cite{Colferai:2010wu}.
Our results show a clear tendency for
$\langle \cos\phi \rangle$ to decrease with $Y$,
whereas~\cite{Colferai:2010wu} predicts a flat $Y$-dependence of
$\langle \cos\phi \rangle$. A possible explanation of this discrepancy, as
pointed our in a very recent paper~\cite{recent}, could be the different
treatment of next-to-NLA corrections, which are beyond the precision of both
studies.

Instead, for the observable $\langle \cos(2\phi) \rangle$ the agreement
between two approaches turns to be rather good, as sees in
Table~\ref{tab:comp_Colferai}, sixth and seventh columns.

Similarly, we compare our predictions for $C_0$ at the present LHC energy
$\sqrt s=7$ TeV with those of Ref.~\cite{Ducloue:2012bm}. The agreement is
 rather  fair, with a discrepancy of about $10\%$, except for the
case $Y=10$, as shown in Table~\ref{tab:comp_Ducloue'}.

\begin{table}[h]
\begin{center}
\begin{tabular}{|r|c|c||c|c||c|c||}
\hline
$Y$ & $C_0^{(\rm NLA)}$             & $C_0^{(\rm NLA)}$
& $\langle\cos\phi\rangle^{\rm{NLA}}$
& $\langle\cos\phi\rangle^{\rm{NLA}}$
& $\langle\cos(2\phi)\rangle^{\rm{NLA}}$
& $\langle\cos(2\phi)\rangle^{\rm{NLA}}$ \\
    & [Ref.~\cite{Colferai:2010wu}] & [here]
    & [Ref.~\cite{Colferai:2010wu}] & [here]
    & [Ref.~\cite{Colferai:2010wu}] & [here]       \\
\hline
6  & 0.606   & 0.726   & 0.851 & 0.763  & 0.512 & 0.457\\
7  & 0.670   & 0.79    & & & & \\
8  & 0.289   & 0.335   & 0.777 & 0.645  & 0.383 & 0.359\\
9  & 0.0474  & 0.0547  & & & & \\
10 & 0.00238 & 0.00272 & 0.753 & 0.574  & 0.317 & 0.285\\
\hline
\end{tabular}
\end{center}
\caption{Comparison of our predictions for the observables $C_0$,
$\langle\cos\phi\rangle$ and $\langle\cos(2\phi)\rangle$ in the NLA for
$|\vec k_{J_1}|=|\vec k_{J_2}|=35$ GeV at $\sqrt{s}=14$ TeV
with those of Ref.~\cite{Colferai:2010wu}.}
\label{tab:comp_Colferai}
\end{table}

\begin{table}[h]
\begin{center}
\begin{tabular}{|r|c|c|c|c|c|c|}
\hline
$Y$ & $C_0^{(\rm NLA)}$ & $\langle\cos\phi\rangle^{\rm{NLA}}$
& $\langle\cos(2\phi)\rangle^{\rm{NLA}}$ & $Y_0$ & $n_R$ \\
\hline
6  & 0.651       & 0.845    & 0.510 & 0 & 1 \\
7  & 0.650       &          &       & 0 & 1 \\
8  & 0.228       & 0.967    & 0.472 & 0 & 1 \\
9  & 0.025       &          &       & 0 & 1 \\
10 & $-$0.000215 &   -      &   -   & 0 & 1 \\
\hline
\end{tabular}
\end{center}
\caption{Values of $C_0$, $\langle\cos\phi\rangle$ and
$\langle\cos(2\phi)\rangle$ in the NLA for $|\vec k_{J_1}|=|\vec k_{J_2}|=35$
GeV at $\sqrt{s}=14$ TeV for the kinematic values of the scales,
$\mu_R^2=s_0=|\vec k_{J_1}||\vec k_{J_2}|$ or $Y_0=0$ and $n_R=1$.}
\label{tab:kinematic}
\end{table}

\begin{table}[h]
\begin{center}
\begin{tabular}{|r|c|c|c|}
\hline
$Y$ & $C_0^{(\rm NLA)}$ & $C_0^{(\rm NLA)}$ \\
    & [Ref.~\cite{Ducloue:2012bm}]  & [here] \\
\hline
6  & 0.172       & 0.186       \\
7  & 0.135       & 0.149       \\
8  & 0.0220      & 0.0247      \\
9  & 0.0007502   & 0.00088     \\
10 & 1.216\ $10^{-6}$ & 5.73\ $10^{-7}$ \\
\hline
\end{tabular}
\end{center}
\caption{Comparison of our predictions for the observable $C_0$
for $|\vec k_{J_1}|=|\vec k_{J_2}|=35$ GeV at $\sqrt{s}=7$ TeV
with those of Ref.~\cite{Ducloue:2012bm}.}
\label{tab:comp_Ducloue'}
\end{table}

Traditionally the BFKL predictions are assumed to be compared with the fixed
order DGLAP
ones, trying to find a kinematic range where possible experiment can
discriminate between these two approaches. For Mueller-Navelet process
the relevant parameter which can describe the separation of these two
regimes is $\eta=\bar \alpha_s(|\vec k_{J}|) Y$, which has the meaning of
the mean number of hard undetected partons inclusively produced in the
process.  For $|\vec k_{J}|=35$ GeV and $Y=6\div 10$  this parameter takes
the values $\eta=0.82\div 1.37$. In Section~3 we
presented also our predictions for smaller jets transverse momenta,
$|\vec k_{J}|=20$ GeV, where $\eta=0.92\div 1.53$ for $Y=6\div 10$.
In this case the
BFKL description is expected to give results more different with respect
to the NLO DGLAP ones. We hope that experiments with such Mueller-Navelet
jet transverse momenta will be possible in the future at LHC.

Our PMS optimization procedure for the kinematics considered gives us optimal
energy and factorization scales values which are substantially larger than the
scale given by the kinematics, $\mu_R^2=|\vec k_{J_1}||\vec k_{J_2}|$,
 especially for higher values of $Y$. This
fact indicates the presence of important contributions subleading to the NLA.
Therefore the estimates of the uncertainties in our predictions should be
taken with care. Nevertheless they reflect the reliability of PMS method in
the considered kinematics. Note that despite very large negative contributions
to the MN-jet cross section coming both from the NLA corrections to the BFKL
kernel and NLA corrections to the jet vertices with respect to LLA MN-jet
cross section, we got with our PMS procedure rather precise results in all
cases.
We point out that this approach was unsatisfactory in the case of asymmetric
kinematics, as already happened in previous studies~\cite{mesons}; in this
kinematics it could be noteworthy to turn to approaches based on collinear
improvement of the BFKL kernel~\cite{Salam:1998tj,Vera:2005jt}. 

It would be very interesting to confront our predictions with experiment.
However, we must  mention that the numerical results we presented here
 refer to  cross sections and angular decorrelations that are differential
in the jet transverse momenta, whereas experimental data would inevitably
include some bins in the transverse momenta of the jets.
For  wide bins in the jet transverse momentum, one definitely needs to
perform a new numerical calculation with our formulas, but this goes beyond
the scope of our present study.

At the very end let us comment a situation related with the kinematics when
jets are basically back-to-back in the transverse plane (with transverse
momenta set equal $|\vec k_{J_1}|=|\vec k_{J_2}|$ or very close to each other).
It has been known for some time~\cite{Frixione:1997ks} that in this kinematic
range the standard NLO QCD collinear calculations for dijet production in
electron-proton collisions exhibit some unreliable features. In particular,
the NLO inclusive dijet total cross section for $ |\vec k_{J_1}|> E_{\rm{cut}}$
and $|\vec k_{J_2}|> E_{\rm{cut}}+\Delta$, contrary to expectations based on
simple phase-space considerations, does not exhibit a monotonic decrease for
increasing $\Delta$, but has instead a local maximum at some small values of
$\Delta$, $\Delta_{\rm{max}}( E_{\rm{cut}})\ll  E_{\rm{cut}}$. A similar
feature of QCD NLO prediction was observed also for the production of dijets
in proton-proton collisions in the case when a (nearly) symmetric restriction
on the transverse momenta of the two jets is imposed,
see~\cite{Andersen:2001kta,Alioli:2010xa}.
The qualitative explanation of such a failure of the NLO QCD approach
(the dip in the cross section at $\Delta\to 0$) was suggested by the authors
of~\cite{Frixione:1997ks} and is related to the fact that, by imposing on the
two  tagged  jets back-to-back condition in the transverse momentum plane,
one effectively puts a veto on the emission of a soft real gluon. This veto
prevents the compensation of the large negative contribution coming from the
soft-virtual term. In this case, an all-order resummation of soft-gluon
effects is needed to get a consistent result. Such resummation was considered
for electron-proton collision in~\cite{Banfi:2003jj} and it is incorporated
in the POWHEG method~\cite{Nason:2004rx,Frixione:2007vw}. Having soft gluon
resummation built in through the parton showering, the POWHEG predictions for
dijet production in proton-proton collisions exhibit a monotonic decrease with
$\Delta$, in contrast to the NLO case, see~\cite{Alioli:2010xa}.

In the case of Mueller-Navelet jet production the situation with the
above-mentioned  symmetric transverse momentum jet tagging
is different from the inclusive dijet total cross section. In Mueller-Navelet
jet kinematics the  tagged  jets are separated for sure by a large
rapidity interval $Y$, which reduces the price paid for the radiation of
hard unidentified partons. Roughly speaking, each additional hard-parton
radiation ``costs'' a factor $\sim \eta=\bar \alpha_s(|\vec k_{J}|) Y$, being
of order of unity for our kinematics.
Note that after a single additional hard-parton emission the transverse momentum
deposit in the event is essentially redistributed and the symmetric transverse
momentum jet tagging condition ($|\vec k_{J_1}|> E_{\rm{cut}}$
and $|\vec k_{J_2}|> E_{\rm{cut}}$) cannot play anymore the role of a veto for
the subsequent emission of real gluons. In this case the compensation of
large effects coming from the virtual and real emission of soft gluons is
effective and one expects a small net effect in the sum of real and virtual
soft gluon contributions.

\section*{Acknowledgements}

We are grateful to the authors of Ref.~\cite{Ducloue:2012bm} for providing
us with the data reported in the second column of
Table~\ref{tab:comp_Ducloue'}. We acknowledge also some stimulating
discussions with Agustin Sabio Vera.

D.I. thanks the Dipartimento di Fisica dell'U\-ni\-ver\-si\-t\`a della Calabria
and the Istituto Nazio\-na\-le di Fisica Nucleare (INFN), Gruppo collegato di
Cosenza, for the warm hospitality and the financial support. The work
of D.I.  was also supported in part by the grants and RFBR-11-02-00242 and
NSh-3810.2010.2. The work of F.C. was supported by European Commission, European
Social Fund and Calabria Region, that disclaim any liability for the use
that can be done of the information provided  in this paper.

\end{document}